\documentclass[10pt, conference]{IEEEtran}
\IEEEoverridecommandlockouts

\usepackage{amssymb}
\usepackage{booktabs}
\usepackage{multirow} 
\usepackage{url}
\usepackage{pifont}
\usepackage{graphicx} %
\usepackage{makecell}
\usepackage{colortbl}  
\usepackage{xcolor}    
\definecolor{lb}{HTML}{D8F3FD} 
\usepackage{enumitem}
\usepackage{mdframed}
\usepackage{algorithm}
\usepackage{algpseudocode}
\usepackage{algorithmicx}
\definecolor{red}{HTML}{E60012} 
\definecolor{green}{HTML}{779649} 
\usepackage{cite}
\usepackage{adjustbox}
\usepackage[T1]{fontenc}
\usepackage{balance}
\usepackage{xspace}
\usepackage{listings}

\definecolor{lightlinkcolor}{HTML}{005595} 
\newcommand{\appendixurl}{\href{https://zenodo.org/records/15561104}{Appendix}}
\newcommand{\appname}{{\textsc{CoSIL}}\xspace}
\newcommand{\appnamebold}{\textbf{\textsc{CoSIL}}\xspace}
\newcommand{\appnameblank}{{\textsc{CoSIL}}}

\usepackage{hyperref}
\hypersetup{colorlinks = true,
	linkcolor = black,
	urlcolor = lightlinkcolor,
	citecolor = black} 

\def\BibTeX{{\rm B\kern-.05em{\sc i\kern-.025em b}\kern-.08em
    T\kern-.1667em\lower.7ex\hbox{E}\kern-.125emX}}

\newcounter{rq}  %
\newcommand{\answerRQ}[1]{\refstepcounter{rq}
\begin{mdframed}[linecolor=gray,roundcorner=12pt,backgroundcolor=gray!15,linewidth=3pt,innerleftmargin=2pt, 
leftmargin=0cm, rightmargin=0cm, topline=false, bottomline=false, rightline=false]
\textbf{Answer to RQ\arabic{rq}:} #1
\end{mdframed}
}

\newcommand{\lzx}[1]{{\it\color{blue}}}
\newcommand{\todo}[1]{{\it\color{red}}}

\begin{document}

\title{Issue Localization via LLM-Driven\\Iterative Code Graph Searching}

\author{%
  \IEEEauthorblockN{%
    Zhonghao Jiang\IEEEauthorrefmark{2},\,
    Xiaoxue Ren\IEEEauthorrefmark{2},\,
    Meng Yan\IEEEauthorrefmark{3},\,
    Wei Jiang\IEEEauthorrefmark{4},\,
    Yong Li\IEEEauthorrefmark{4},\,
    Zhongxin Liu\IEEEauthorrefmark{2}\IEEEauthorrefmark{1}%
  }%
  \vspace{1ex}%
  \IEEEauthorblockA{\IEEEauthorrefmark{2}The State Key Laboratory of Blockchain and Data Security, Zhejiang University, Hangzhou, China\\
  }
  \IEEEauthorblockA{\IEEEauthorrefmark{3}School of Big Data and Software Engineering, Chongqing University, Chongqing, China\\
  }
  \IEEEauthorblockA{\IEEEauthorrefmark{4}Ant Group, Hangzhou, China\\
  Emails: \{zhonghao.j, xxren, liu\_zx\}@zju.edu.cn, mengy@cqu.edu.cn, \{jonny.jw, liyong.liy\}@antgroup.com
  }
  \thanks{\IEEEauthorrefmark{1} Corresponding author.}
}

\maketitle

\begin{abstract}

Issue solving aims to generate patches to fix reported issues in real-world code repositories according to issue descriptions.
Issue localization forms the basis for accurate issue solving.
Recently, large language model (LLM) based issue localization methods have demonstrated state-of-the-art performance.
However, these methods either search from files mentioned in issue descriptions or in the whole repository and struggle to balance the breadth and depth of the search space to converge on the target efficiently.
Moreover, they allow LLM to explore whole repositories freely, making it challenging to control the search direction to prevent the LLM from searching for incorrect targets.
Meanwhile, because LLMs may not correctly produce the required interaction formats with the environment, they suffer from search failures.

This paper introduces \appnamebold, an LLM-driven, powerful function-level issue localization method without training or indexing. 
To balance search breadth and depth, \appnamebold employs a two-phase code graph search strategy.
It first conducts broad exploration at the file level using dynamically constructed module call graphs, and then performs in-depth analysis at the function level by expanding the module call graph into a function call graph and executing iterative searches.
To precisely control the search direction, \appnamebold designs a pruner to filter unrelated directions and irrelevant contexts. 
To avoid incorrect interaction formats in long contexts, \appnamebold introduces a reflection mechanism that uses additional independent queries in short contexts to enhance formatted abilities.
Experiment results demonstrate that \appnamebold achieves a Top-1 localization accuracy of 43.3\% and 44.6\% on SWE-bench Lite and SWE-bench Verified, respectively, with Qwen2.5-Coder-32B, average outperforming the state-of-the-art methods by 96.04\%. 
When \appnamebold is integrated into an issue-solving method, Agentless, the issue resolution rate improves by 2.98\%–30.5\%.

\end{abstract}

\begin{figure*}[th!]
    \centering
    \includegraphics[width=0.8\linewidth]{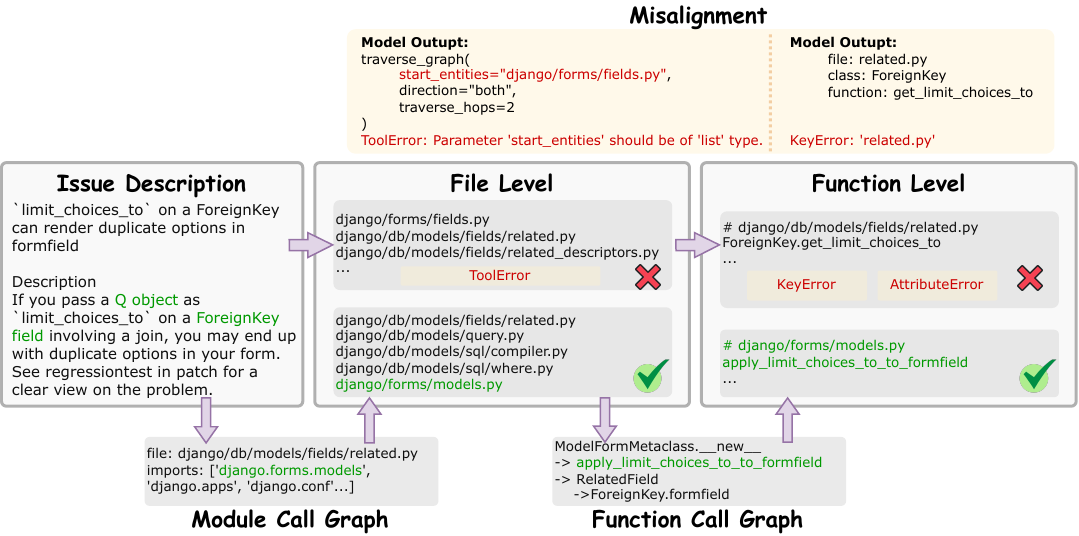}
    \caption{A motivating example of Django-13315.}
    \label{motivation example}
\end{figure*}

\section{Introduction}

Recently, issue solving~\cite{swebench,swebenchm} has become a crucial task as it directly addresses bugs \cite{zhao2024enhancing, yin2024thinkrepair, bouzenia2025repairagent} and feature requests \cite{li2025feabenchbenchmarkevaluatingrepositorylevel, deng2025nocode} in real-world code repositories, thereby improving software quality and ensuring the continuous evolution of repositories.
Given an issue in a repository, issue-solving methods aim to generate a patch according to the issue description provided by users.
Accurate localization is the foundation for successfully resolving issues~\cite{meng2024empiricalstudyllmbasedagents}. 
It aims to search for a set of suspicious code snippets within a certain scope of code files (i.e., search space) of the code repository.

LLM-based methods \cite{agentless, orcaloca, locagent, chang2025bridging} have become the main force in issue localization tasks.
Existing LLM-based issue localization methods can be mainly divided into prompt-based methods and agent-based methods.
They face the following common challenges.
\ding{182} \textbf{They fail to balance the breadth and depth of the search space.}
Prompt-based methods~\cite{agentless, repograph} use the issue description and employ a hierarchical strategy to gradually converge to the target. 
However, issue description usually describes the symptoms rather than the essence of the problem~\cite{chen2025unveiling}, resulting in an overly narrow search space and causing relevant context to be ignored.
Agent-based methods~\cite{orcaloca,locagent,wang2024openhands} leverage the code understanding capability of LLMs to freely explore the entire repository, which results in an overly broad search space and leads to the retrieval of irrelevant context. 
Irrelevant context interferes with useful information, leading to performance degradation.
\ding{183} \textbf{They lack control of the search direction.}
Prompt-based methods often use as much context as possible from the search space without verifying its relevance. 
This causes the LLM to explore some search directions yet unrelated to the issue, making it difficult to locate the correct target.
Agent-based methods directly depend on agents to decide search direction, which often rely only on the current state and lack a global perspective~\cite{liu2024large}.
As a result, the agent may repeatedly explore the same code segments, leading to chaotic search directions. 
Additionally, as the agent bases its subsequent decisions on historical feedback from the environment, chaotic search directions can propagate, resulting in redundant noise in the context and degrading the performance of issue localization.
\ding{184} \textbf{They produce abnormal search processes due to incorrectly formatted interaction outputs.}
LLM-based methods interact with the external environment (e.g., tool calling) via formatted outputs from the LLM~\cite{wang2024openhands, yang2024sweagent}.
Prompt-based methods rely entirely on the LLM’s inherent capabilities, which directly treat outputs with incorrectly formatted interactions as localization failures~\cite{agentless,repograph}, resulting in empty lists in the localization results.
Agent-based methods typically add retry mechanisms, but the error messages produced by incorrectly formatted interactions are also included in the context~\cite{locagent}. 
As consecutive retries fill the context with error messages, they interfere with localization performance.
Thus, the limited capability of LLMs to follow instructions for formatted output in long-context settings~\cite{bai2024longalignrecipelongcontext} becomes a performance bottleneck.

To overcome the challenges above, we propose \appname, an LLM-driven issue localization technique through iterative code graph searching. 
\textbf{To balance the breadth and depth of the search space}, we adopt a two-stage search strategy that performs broad exploration at the file level and in-depth analysis at the function level.
Specifically, we provide the LLM with a module call graph to expand the search space beyond the files mentioned in the issue description, enabling the identification of additional suspicious modules.
By further expanding the module call graph into a function call graph and performing an iterative search, we collect highly relevant contextual information in greater depth.
\textbf{To control the search direction}, we design a pruner that rejects the LLM's exploration of irrelevant context.
This ensures the effectiveness of the collected contextual information and filters out contextual noise.
Given LLMs exhibit stronger instruction-following capabilities in short-context settings~\cite{bai2024longalignrecipelongcontext}, we leverage a reflection mechanism \cite{shinn2023reflexionlanguageagentsverbal}  that summarizes decision information into a short context \textbf{to prevent the abnormal interaction format through enhancing LLM's formatted outputs}.
It involves performing an additional, independent query at the end of each LLM interaction to double-check and format the final output.

We implement \appname based on Qwen2.5-Coder-7B/14B/32B \cite{hui2024qwen2}, and evaluate its effectiveness on SWE-bench Lite \cite{swebench} and SWE-bench Verified \cite{swebenchverified}. 
Experimental results show that, compared to existing LLM-based issue localization methods, \appname average improves Top-1 localization accuracy by 6.61\% at the file level and by 96.04\% at the function level.
Furthermore, when leveraging the localization results generated by \appname is integrated into an issue-solving method, Agentless~\cite{agentless}, the issue resolution rate increases by 2.98\%–30.5\%, further demonstrating the usefulness in application scenarios.
In addition, we conduct an ablation study on the key components of \appname, i.e., reflective alignment, module call graph, iterative search, and pruning.
The results demonstrate the effectiveness of \appname's key components: iterative search as the core, call graphs for search space constraint, pruning for search direction management, and reflective alignment for ensuring interaction format.
Finally, we evaluate \appname on different families of LLMs and compute its cost.
Results show that it can generalize with a cost of 0.02\$ and 0.34\$ on DeepSeek-v3-0324~\cite{guo2025deepseek} and GPT-4o-2024-0806~\cite{openai2024gpt4ocard}.

In summary, our main contributions are as follows:
\begin{itemize}[left=0pt, topsep=0em]
    \item We propose \appname, an LLM-driven function-level issue localization method without training or pre-indexing. 
    It balances the breadth and depth of the search space while ensuring the accuracy of the search direction and the correct interaction formats.
    \item We conduct extensive and comprehensive experiments on SWE-bench Lite \cite{swebench} and SWE-bench Verified \cite{swebenchverified}.
    Experiment results show that \appname achieves up to 48\% Top-1 accuracy in function-level localization, nearly twice that of existing methods. 
    It brings an average performance improvement of 14.90\% on the issue-solving task.
    \item We open-source the replication package, including the source code and data of \appname, at \url{https://github.com/ZJU-CTAG/CoSIL}. 
\end{itemize}

\begin{figure*}[t!]
    \centering
    \includegraphics[width=0.95\linewidth]{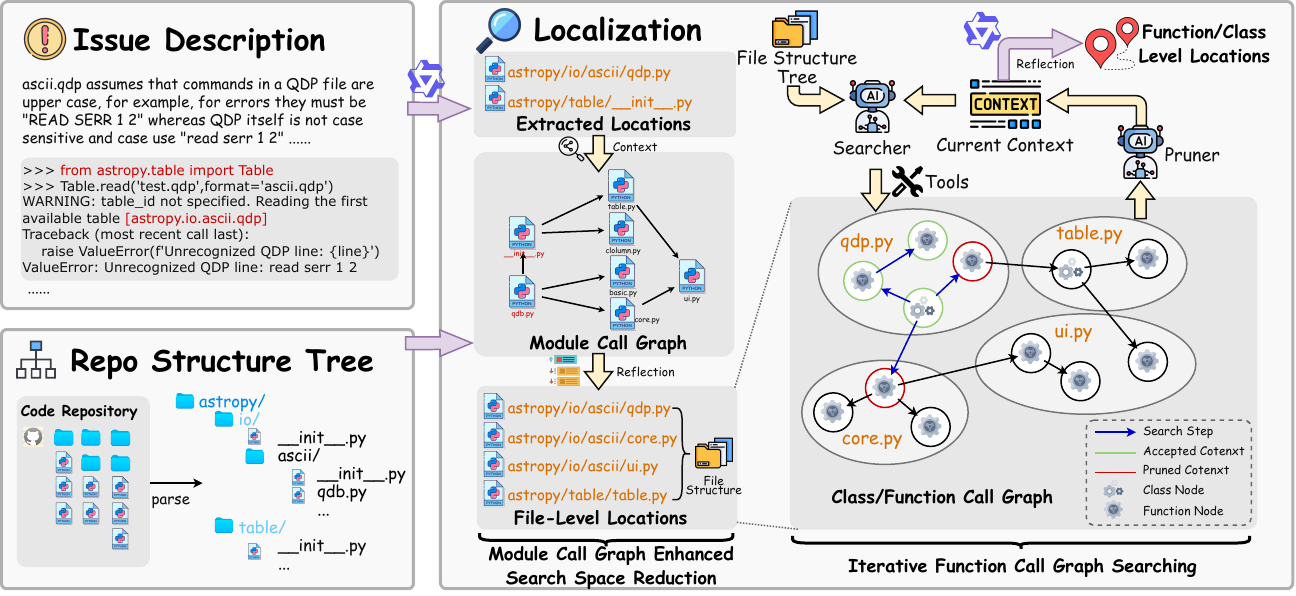}
    \caption{Overview of \appname.}
    \label{overview}
\end{figure*}

\section{Motivation and Background}

\subsection{Motivating Example}
Figure~\ref{motivation example} presents a motivating example, which is an issue collected from the Django repository.

This example describes a phenomenon where passing a ``\emph{Q Object}'' to the ``\emph{limit\_choices\_to}'' parameter of a ``\emph{ForeignKey Field}'' results in duplicate options appearing in the form.
We inspect the results produced by three state-of-the-art issue localization methods, i.e., Agentless \cite{agentless}, OrcaLoca \cite{orcaloca}, and LocAgent \cite{locagent}, using Qwen2.5-Coder-32B model. 
None of them successfully find the correct location to edit.

We analyze their trajectories when handling this case and summarize three major observations.
\ding{182} The three methods are limited in search breadth or depth. Specifically, Agentless extracts targeted file names mentioned in the issue description \cite{agentless}, including ``\path{django/forms/fields.py}'' and ``\path{django/db/models/fields/related.py}''.
While these files are related to the issue, they are not the ones that require modification. 
Conducting function-level localization within such a narrow search space makes it nearly impossible to identify the correct function to edit.
In contrast, OrcaLoca searches by analyzing the frequency with which the LLM focuses on each function~\cite{orcaloca}.
But the most frequently referenced function, ``\emph{ForeignKey.get\_limit\_choices\_to}'', may only be textually similar to the issue description but semantically irrelevant, searching in such an overly broad space as the entire repository is not feasible.
\ding{183} These methods lack control of search directions.
Specifically, when LocAgent uses the ``\emph{search\_code\_snippets}'' tool to read code snippets, both the first and third actions inspect the source code of ``\emph{ForeignKey}'', resulting not only in redundant context but also in an unordered search process. 
\ding{184} Incorrectly formatted outputs during interactions lead to the three methods crashing.
Specifically, when LocAgent \cite{locagent} calls the ``\emph{traverse\_graph}'' tool, where the LLM passes an argument with an incorrect type for the parameter ``\emph{start\_entities}'', it not only fails to retrieve context, but also includes detailed error messages of this ``\emph{ToolError}''.
Moreover, when these three methods feed back the decided locations, the localization may fail because unformatted output cannot be parsed.
For example, Agentless \cite{agentless} might output an incomplete file path, leading to a ``\emph{KeyError}'' and ultimately a failed localization attempt.

From observation 1, we find that when the search space is too narrow, existing techniques cannot find the target within a space that excludes the correct one.
When the search space is too broad, existing techniques may retrieve irrelevant targets that are only textually similar.
This motivates a two-phase strategy to balance the search space, i.e., expanding the search breadth at the file level through the module call graph while maintaining search depth through function call graphs.
From observation 2, we find that if the search direction is not restricted, the limited context window becomes populated with substantial redundant or unrelated content, leading to inefficiency. 
This motivates us to prune the irrelevant context to control the search directions.
From observation 3, we find that incorrect output format during the LLM's interaction with the environment not only pollutes the context with irrelevant information but can even derail the localization process entirely.
Motivated by the fact that LLM alignment degrades in long-context settings \cite{bai2024longalignrecipelongcontext, zhang2025sealignalignmenttrainingsoftware}, we integrate self-reflection \cite{shinn2023reflexionlanguageagentsverbal} as a verifier to double-check and correct final decisions.

\subsection{Call Graph Construction}\label{codegraph}

As the call graphs in the repository typically describe the invocation and dependency relationships among modules at the class/function level, we can use them to expand the search space at a coarse granularity and to constrain the search space at a fine granularity.

In this work, we mainly use two types of call graphs: module call graphs and function call graphs.
Given a code repository \( R \), the module call graph is defined as \( \mathcal{G}_M(R) = (\mathcal{V}_m, \mathcal{E}_m) \), where \( \mathcal{V}_m \) consists of all modules in \( R \), and \( \mathcal{E}_m \) includes all ``import'' relationships between modules.
Furthermore, each node in \( \mathcal{V}_m \) can be expanded into multiple class/function nodes. Based on this, the function call graph can be defined as \( \mathcal{G}_F(R) = (\mathcal{V}_f, \mathcal{E}_f) \), where \( \mathcal{V}_f \) consists of all the classes and functions defined in \( \mathcal{V}_m \). \( \mathcal{E}_f \) includes all ``invoke'' and ``inherit'' relationships between classes and functions.
We use the LLM to construct the necessary call graph dynamically in a textual representation. 
Specifically, we provide the LLM with the code of the module/function and its import statements parsed from the files where they are located, prompting the LLM to analyze their outgoing dependency and generate a first-order subgraph centered on the target module/function.
\textbf{An example of the call graph and the detailed construction algorithm through LLMs can be found in our online \appendixurl-A.}

\section{Approach}

\begin{figure}[t!]
    \centering
    \includegraphics[width=1\linewidth]{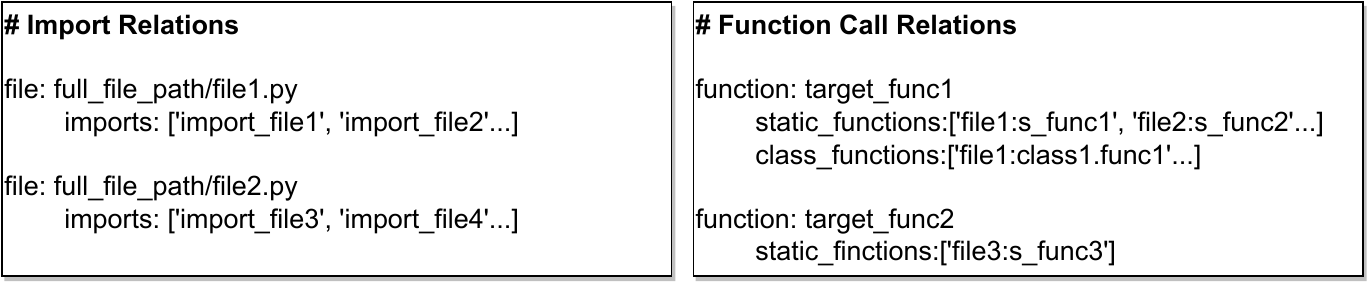}
    \caption{Textual representation of call graphs.}
    \label{textualgraph}
\end{figure}

In this section, we introduce \appname, an LLM-driven issue localization framework. 
As shown in Figure \ref{overview}, \appname takes an issue description and a repository structure tree extracted from the original code repository as inputs and returns a recommendation list containing the signatures of the most suspicious functions. 
It consists of two stages, i.e., module call graph enhanced search space reduction at the file level and iterative function call graph searching at the function level.

To ensure that the search space is not limited and avoid an overly broad scope that includes the entire repository, \appname expands the search space from the modules mentioned in the issue description to imported modules by providing the LLM with the module call graph during the file-level stage (Section \ref{step1}). 
To collect context and perform an in-depth search, \appname uses a specifically designed search tool to iteratively explore the function call graph at the function-level stage (Section \ref{step2}). 
Additionally, to achieve a balance between context length and the density of relevant information, \appname utilizes a pruning mechanism to control the search direction and prevent irrelevant context from being retrieved (Section \ref{step3}).
To enhance the LLM’s ability to produce formatted outputs when interacting with the external environment,  \appname leverages reflection at the end of each stage, thereby alleviating parsing errors in the final results (Section \ref{step0}).

\subsection{Localization with Call Graph}
\subsubsection{Module Call Graph Enhanced Search Space Reduction}\label{step1}

This stage takes the repository structure tree as input and returns a list of suspicious files as the search space.
Because of the limited context window size of LLMs, directly providing the code in all files to the model is often infeasible. 
Inspired by previous studies \cite{agentless, repograph}, we recursively traverse the entire repository and parse it into a hierarchical repository-structure tree representation.
Considering that most issue reports contain information about candidate suspicious locations \cite{zhang2024autocoderover}, we query LLM with the issue description to pre-select the related files
The prompt template can be found in \appendixurl-E. 
Then, the ``\textit{import}'' statements in the related files are parsed to construct a first-order subgraph of the module call graph, which starts from the related files.
Such a subgraph is then converted into structured textual representations, as shown in Figure \ref{textualgraph}.
Next, the pre-selected related files, along with the repository-structure tree and the module call graph, are fed into the LLM for search space expansion, enabling the reselection and reranking of the suspicious files. 
The rationale behind not utilizing the entire module call graph is that the repository-structure tree already provides a comprehensive view of the repository.

\begin{table}[t]
\centering
\caption{Design of search tools.}
\resizebox{\columnwidth}{!}{
\begin{tabular}{ll}
\toprule
\textbf{Function Name} & \textbf{Description} \\
\midrule
search\_class\_node & Get the code snippet of a class node.\\
search\_class\_function\_node & Get the code snippet of a class member function node.\\
search\_file\_function\_node & Get the code snippet of a static function node.\\
exit& Terminate iteration search and go into summary phase.\\
\bottomrule
\end{tabular}
}
\label{tools}
\end{table}

\subsubsection{Iterative Function Call Graph Searching}\label{step2}
This stage takes the file-level search space as input and returns a list of suspicious functions.
Specifically, it consists of the following three steps. (\textbf{Detailed algorithm is in \appendixurl-B})

\textbf{Step I: Initialize the search state.} 
This step aims to preliminarily determine one or several starting points for the search.
Specifically, similar to the repository-structure tree described in Section \ref{step1}, we construct a file-structure tree for all suspicious files, outlining all classes, member functions, and static functions within these files in a hierarchical format. 
Then we provide the issue description along with this file-structure tree as inputs to query the LLM to return selected points.

\textbf{Step II: Iterative search contexts.} 
This step aims to collect contextual information for locating suspicious functions.
First, we construct a search agent by equipping the LLM with retrieval tools as shown in Table \ref{tools} and a pruner agent (detailed in Section \ref{step3}) by providing a function code snippet and asking for a boolean flag indicating if the code snippet should be used as context.
Then, we set a maximum number of iterations to prevent the search agent from falling into an infinite loop.
During each iteration step, the search agent first selects a target node from the set of accessible nodes (denoted as $visableNodes$) for exploration according to the issue description and the visited nodes (denoted as $contextNodes$).
The target node could refer to a class, a class member function, or a static function.
Depending on the type of the target node, the search agent invokes different search tools shown in Table \ref{tools} to retrieve node content.
Next, the retrieved code will be verified by the pruner agent, which determines whether the target node is acceptable.
Once the target node is confirmed and accepted by the pruner agent into $contextNodes$, its neighbor nodes within the function call graph become accessible and are added to the $visableNodes$ for selection in subsequent iterations.
Notably, if the search agent returns an incorrect function-call parameter, i.e., the search agent would like to retrieve non-existent code elements, a post-processing procedure will prompt the search agent to reselect the node from the accessible nodes for exploration.
Additionally, if the search agent invokes the "exit" tool, which means that sufficient information has been collected to support a decision, the iteration process will be terminated prematurely.

\textbf{Step III: Summarize the suspicious functions.}
This step aims to select and rerank a set of the most suspicious functions or classes from the context nodes generated by Step II.
Specifically, the search agent incorporates the retrieved code snippets from relevant context nodes into a summarization prompt as the background knowledge for in-context learning.
Then, LLM returns a list of suspicious functions as localization results.
The summarized function-level localization information can either be utilized for further fine-grained line-level localization or provided as contextual input to the LLM for subsequent program repair directly.

\subsection{Context Management with Pruning}\label{step3}
This component takes the issue description and code snippet to be explored by the search agent as input and returns a boolean flag indicating whether the node should be accepted.
Specifically, when LLM selects a target node, the code snippet of the node is used as context to construct a prompt that instructs the LLM to analyze whether the node is related to the issue being addressed, and the LLM is asked to return a boolean value. 
If the pruner agent returns a ``True'' flag to accept this node as part of the context, its neighboring nodes will subsequently be expanded. 
Otherwise, the pruner directly rejects the expansion and excludes the node from the context, effectively pruning the search path.

By pruning the search path, \appname effectively guides the search direction, encouraging the LLM to explore suspicious functions to avoid getting stuck in local optima.
Simultaneously, it enables efficient context management, preventing the agent from collecting excessive redundant information, which might lead to erroneous decision-making.

\subsection{Reflective Alignment}\label{step0}
This component takes the last returned response by the LLM in Section~\ref{step1} and Section~\ref{step2} to rerank and reformat the output.
It acts as a verifier to guard against incorrect interaction formats in long-context scenarios~\cite{bai2024longalignrecipelongcontext}.

Specifically, we maintain a message list to record each query and response during interactions with the LLM, and collect context through multiple queries. 
Since \appname requires the LLM to summarize the suspicious locations in the final query round (Section~\ref{step1} and~\ref{step2}), the last response in the message list should contain the LLM’s decision information.
Reflective Alignment takes the issue description together with the last response containing decision information as context, and prompts the LLM to rerank and reformat the candidate locations given in the decision information to obtain the final localization result.
The detailed prompt can be found in \appendixurl-E.
Reflective Alignment is applied at the end of both file-level and function-level localization, significantly reducing localization failures caused by unstructured or improperly formatted outputs.
It also retains the advantages of the reflection reasoning strategy \cite{shinn2023reflexionlanguageagentsverbal}, and thus can correct and rerank the localization results.

\begin{table*}[!ht]
\centering
\caption{Localization results under different LLMs on SWE-bench Lite. ER indicates \emph{empty rate}.}
\begin{adjustbox}{width=\textwidth}
\begin{tabular}{cccccc|ccccc|c}
\toprule
\multirow{2}{*}{\textbf{Method}} & \multicolumn{5}{c}{\textbf{File-level}}  & \multicolumn{5}{c}{\textbf{Function-level}} &  \\
\cmidrule{2-12}
& \textbf{Top-1}   & \textbf{Top-3}   & \textbf{Top-5} & \textbf{MAP} & \textbf{MRR}  & \textbf{Top-1}   & \textbf{Top-3}   & \textbf{Top-5}  & \textbf{MAP} & \textbf{MRR} & \textbf{ER} \\
\midrule
\multicolumn{12}{c}{\cellcolor{lb} Qwen2.5-Coder-7B} \\
\midrule
Agentless-FL & 0.333~\textcolor{red}{↑23.1\%} & 0.537~\textcolor{red}{↑7.4\%} & 0.577~\textcolor{red}{↑10.4\%} & 0.437~\textcolor{red}{↑20.6\%} & 0.437~\textcolor{red}{↑14.4\%} & 0.173~\textcolor{red}{↑23.1\%} & 0.260~\textcolor{red}{↑12.7\%} & 0.270~\textcolor{red}{↑13.7\%} & 0.218~\textcolor{red}{↑10.6\%} & 0.218~\textcolor{red}{↑14.7\%} & 5.33\% \\
OrcaLoca & \textbf{0.467}~\textcolor{green}{↓12.2\%} & 0.490~\textcolor{red}{↑17.7\%} & 0.490~\textcolor{red}{↑30.0\%} & 0.478~\textcolor{red}{↑10.3\%} & 0.478~\textcolor{red}{↑4.6\%} & 0.157~\textcolor{red}{↑35.7\%} & 0.273~\textcolor{red}{↑7.3\%} & 0.293~\textcolor{red}{↑4.8\%} & 0.209~\textcolor{red}{↑15.3\%} & 0.214~\textcolor{red}{↑16.8\%} & 8.33\% \\
LocAgent & 0.400~\textcolor{red}{↑2.5\%} & 0.507~\textcolor{red}{↑15.8\%} & 0.533~\textcolor{red}{↑19.5\%} & 0.458~\textcolor{red}{↑15.1\%} & 0.458~\textcolor{red}{↑9.2\%} & 0.047~\textcolor{red}{↑353.2\%} & 0.140~\textcolor{red}{↑109.3\%} & 0.160~\textcolor{red}{↑91.9\%} & 0.091~\textcolor{red}{↑164.8\%} & 0.092~\textcolor{red}{↑171.7\%} & 21.00\% \\
\appnameblank & 0.410 & \textbf{0.577} & \textbf{0.637} & \textbf{0.527} & \textbf{0.500} & \textbf{0.213} & \textbf{0.293} & \textbf{0.307} & \textbf{0.241} & \textbf{0.250} & \textbf{1.00\%} \\
\midrule
\multicolumn{12}{c}{\cellcolor{lb} Qwen2.5-Coder-14B} \\
\midrule
Agentless-FL & 0.537~\textcolor{red}{↑8.6\%} & 0.687~\textcolor{red}{↑6.7\%} & 0.723~\textcolor{red}{↑6.1\%} & 0.618~\textcolor{red}{↑8.1\%} & 0.611~\textcolor{red}{↑7.7\%} & 0.187~\textcolor{red}{↑8.6\%} & 0.253~\textcolor{red}{↑29.2\%} & 0.280~\textcolor{red}{↑26.1\%} & 0.219~\textcolor{red}{↑16.4\%} & 0.224~\textcolor{red}{↑19.2\%} & 1.00\% \\
OrcaLoca & 0.433~\textcolor{red}{↑34.6\%} & 0.477~\textcolor{red}{↑53.7\%} & 0.477~\textcolor{red}{↑60.8\%} & 0.455~\textcolor{red}{↑46.8\%} & 0.455~\textcolor{red}{↑44.6\%} & 0.170~\textcolor{red}{↑19.4\%} & 0.303~\textcolor{red}{↑7.9\%} & 0.313~\textcolor{red}{↑12.8\%} & 0.229~\textcolor{red}{↑11.4\%} & 0.233~\textcolor{red}{↑14.6\%} & 6.33\% \\
LocAgent & 0.510~\textcolor{red}{↑14.3\%} & 0.633~\textcolor{red}{↑15.8\%} & 0.650~\textcolor{red}{↑18.0\%} & 0.572~\textcolor{red}{↑16.8\%} & 0.572~\textcolor{red}{↑15.0\%} & 0.090~\textcolor{red}{↑125.6\%} & 0.283~\textcolor{red}{↑15.5\%} & 0.317~\textcolor{red}{↑11.3\%} & 0.192~\textcolor{red}{↑32.8\%} & 0.190~\textcolor{red}{↑40.5\%} & 22.33\% \\
\appnameblank & \textbf{0.583} & \textbf{0.733} & \textbf{0.767} & \textbf{0.668} & \textbf{0.658} & \textbf{0.203} & \textbf{0.327} & \textbf{0.353} & \textbf{0.255} & \textbf{0.267} & \textbf{0.33\%} \\
\midrule
\multicolumn{12}{c}{\cellcolor{lb} Qwen2.5-Coder-32B} \\
\midrule
Agenltess-FL & 0.583~\textcolor{red}{↑5.1\%} & 0.730~\textcolor{red}{↑6.8\%} & 0.773~\textcolor{red}{↑8.3\%} & 0.669~\textcolor{red}{↑5.4\%} & 0.659~\textcolor{red}{↑6.4\%} & 0.247~\textcolor{red}{↑75.3\%} & 0.397~\textcolor{red}{↑37.8\%} & 0.427~\textcolor{red}{↑35.8\%} & 0.308~\textcolor{red}{↑51.0\%} & 0.325~\textcolor{red}{↑51.7\%} & 2.00\% \\
OrcaLoca & 0.590~\textcolor{red}{↑3.9\%} & 0.640~\textcolor{red}{↑21.9\%} & 0.640~\textcolor{red}{↑30.8\%} & 0.615~\textcolor{red}{↑14.6\%} & 0.615~\textcolor{red}{↑14.0\%} & 0.217~\textcolor{red}{↑99.5\%} & 0.417~\textcolor{red}{↑31.2\%} & 0.447~\textcolor{red}{↑29.8\%} & 0.302~\textcolor{red}{↑54.0\%} & 0.317~\textcolor{red}{↑55.5\%} & 2.67\% \\
LocAgent & \textbf{0.647}~\textcolor{green}{↓5.3\%} & 0.727~\textcolor{red}{↑7.2\%} & 0.740~\textcolor{red}{↑13.1\%} & 0.683~\textcolor{red}{↑3.2\%} & 0.683~\textcolor{red}{↑2.6\%} & 0.103~\textcolor{red}{↑320.4\%} & 0.377~\textcolor{red}{↑45.1\%} & 0.447~\textcolor{red}{↑29.8\%} & 0.231~\textcolor{red}{↑101.3\%} & 0.244~\textcolor{red}{↑102.0\%} & 13.67\% \\
\appnameblank     & 0.613 & \textbf{0.780}& \textbf{0.837}& \textbf{0.705}& \textbf{0.701}& \textbf{0.433}& \textbf{0.547}& \textbf{0.580}& \textbf{0.465}& \textbf{0.493}& \textbf{0.00\%} \\
\bottomrule
\end{tabular}
\end{adjustbox}
\label{effect_lite}
\end{table*}

\section{Experiment Setup}

\subsection{Dataset}
SWE-bench \cite{swebench} is the most widely used dataset for evaluating large language models' capabilities in solving real-world software issues. It extracts 2,294 tasks from 12 Python code repositories, where each task requires submitting a patch to solve an issue in the corresponding repository.
To control experimental costs, we conduct experiments on two popular subsets of SWE-bench:  
\begin{itemize}[left=0pt, topsep=0em]
    \item \textbf{SWE-bench Lite} \cite{swebench} is a 300-instance subset selected using heuristic methods. It removes tasks where the issue description contains images, external hyperlinks, and other non-textual elements. 
    \item \textbf{SWE-bench Verified} \cite{swebenchverified} is a subset of 500 instances developed by OpenAI. It is selected from 1,699 manually annotated instances by professional developers and removes instances with underspecified descriptions or overly specific test cases.
\end{itemize}

\subsection{Baselines}
We mainly compare \appname with open-source localization methods that focus on SWE-bench, including Agentless \cite{agentless}, Orcaloca \cite{orcaloca}, and LocAgent \cite{locagent}.

\begin{itemize}[left=0pt, topsep=0em]
\item \textbf{Agentless} is the current state-of-the-art open-source issue solving pipeline.
It is designed based on human prior knowledge and performs issue localization by hierarchically querying the LLM for suspicious files, functions, and code lines.
We use the localization component in Agentless as the baseline, denoted as \textbf{Agentless-FL}.

\item \textbf{Orcaloca} designs an LLM agent framework, which searches on a unique code graph based on priority scheduling, action decomposition, and relevance scoring. In addition, it employs distance-aware pruning on the retrieved context, enhancing the accuracy of fault localization.

\item \textbf{LocAgent} parses all dependency relationships within the code repository and constructs a large graph index, enabling the LLM to locate relevant entities through multi-hop reasoning using retrieval tools.
\end{itemize}

Additionally, we consider the fine-tuning-based approach BugCerberus \cite{chang2025bridging}, but fail in conducting the comparison because its implementation is not released.

\subsection{Research Questions (RQs)}
To evaluate the performance of \appname under issue localization tasks, we consider the following research questions.

\begin{itemize}[left=0pt, topsep=0em]
    \item \textbf{RQ1. Effectiveness}: How effective is \appname in localizing buggy code snippets in the code repository?

    \item \textbf{RQ2. Ablation}: How do the key components of \appname contribute to its effectiveness?

    \item \textbf{RQ3. Application}: Can the buggy locations identified by \appname lead to better performance in issue solving?

    \item \textbf{RQ4. Generalizability}: How well does \appname generalize across different families of LLMs?
\end{itemize}

\subsection{Evaluation Metrics}
For RQ1 and RQ2, we mainly use the three evaluation metrics, i.e., \emph{Top-N}, \emph{MAP}, and \emph{MRR}. They are widely used to evaluate the performance of fault location methods \cite{flexfl, autofl, agentfl, chang2025bridging}. Additionally, we measure the \emph{empty rate} across the localization results. For RQ3, we use Applied\% and Resolved\% following prior works \cite{yang2024sweagent, zhang2024autocoderover, orcaloca, chang2025bridging}.

\textbf{Top-N:} 
This metric counts how often at least one bug location appears within the top N recommendations. 
To measure the localization success rate, we normalize this count by the number of instances. 
We set N = \{1, 3, 5\}, as approximately 73.58\% of developers consider only the top 5 results \cite{kochhar2016practitioners}.

\textbf{MAP:} Mean Average Precision (MAP) evaluates the ranking quality of buggy elements identified by a technology \cite{autofl, flexfl, agentfl}.

\textbf{MRR:} Mean Reciprocal Rank (MRR) measures the ranking performance of a technology by evaluating the position of the first identified buggy element within the recommendation list~\cite{autofl, flexfl, agentfl}. 

\textbf{Empty Rate}: 
The empty rate refers to the proportion of cases where the suspicious location recommendation list generated by a technique is empty, indicating the technique fails to identify any candidate locations for a given issue.

\textbf{Resolved\% \& Applied\%:}
Resolved\% refers issue resolution rate \cite{swebench}, which measures the percentage of problems that are successfully solved by a technique. Applied\% denotes the application rate, which evaluates the percentage of patches that are generated by a technique and can be successfully applied to the corresponding repositories without syntax errors.

\subsection{Implementation Details}
In our experiments, we deploy the Qwen2.5-coder models (7B, 14B, and 32B) locally using the vLLM framework \cite{vllm2024} following prior studies~\cite{locagent}. Additionally, we employ the official GPT-4o-2024-08-06 and DeepSeek-v3-0324 API to perform supplementary experiments to validate generalizability~\cite{agentless}.

To ensure a fair comparison, we adopt greedy decoding (i.e., sampling with temperature=0) during the localization stage. Moreover, for each instance, we set the maximum number of iterations to 10 during the graph search process. 
For agent-based methods (i.e., OrcaLoca and LocAgent), we set the maximum number of retries per instance to 3 and the maximum search time to 15 minutes, following their default configurations.
When comparing the effectiveness of various localization methods in enhancing issue solving, we uniformly utilized the repair component of Agentless-1.5 \cite{agentless} to generate 10 candidate patches per instance using the predicted top-3 locations. One candidate patch is obtained through greedy decoding, and nine are generated with temperature=0.8. 
Moreover, we use regression tests selected by LLM and LLM-generated reproduction tests to validate the patches following the practice of Agentless-1.5 \cite{agentless}.

\begin{table*}[t!]
\centering
\caption{Localization results under different LLMs on SWE-bench Verified. ER indicates \emph{empty rate}.}
\begin{adjustbox}{width=\textwidth}
\begin{tabular}{cccccc|ccccc|c}
\toprule
\multirow{2}{*}{\textbf{Method}} & \multicolumn{5}{c}{\textbf{File-level}}  & \multicolumn{5}{c}{\textbf{Function-level}} &  \\
\cmidrule{2-12}
& \textbf{Top-1}   & \textbf{Top-3}   & \textbf{Top-5} & \textbf{MAP} & \textbf{MRR}  & \textbf{Top-1}   & \textbf{Top-3}   & \textbf{Top-5}  & \textbf{MAP} & \textbf{MRR} & \textbf{ER} \\
\midrule
\multicolumn{12}{c}{\cellcolor{lb} Qwen2.5-Coder-7B} \\
\midrule
Agentless-FL & 0.514~\textcolor{red}{↑5.8\%} & 0.650~\textcolor{red}{↑7.1\%} & 0.670~\textcolor{red}{↑10.7\%} & 0.551~\textcolor{red}{↑11.1\%} & 0.577~\textcolor{red}{↑8.1\%} & 0.256~\textcolor{red}{↑10.2\%} & 0.306~\textcolor{red}{↑17.0\%} & 0.320~\textcolor{red}{↑15.0\%} & 0.245~\textcolor{red}{↑15.9\%} & 0.282~\textcolor{red}{↑13.1\%} & 8.00\% \\
OrcaLoca & 0.510~\textcolor{red}{↑6.7\%} & 0.534~\textcolor{red}{↑30.3\%} & 0.534~\textcolor{red}{↑39.0\%} & 0.490~\textcolor{red}{↑24.9\%} & 0.522~\textcolor{red}{↑19.5\%} & 0.206~\textcolor{red}{↑36.9\%} & 0.322~\textcolor{red}{↑11.2\%} & 0.340~\textcolor{red}{↑8.2\%} & 0.227~\textcolor{red}{↑25.1\%} & 0.266~\textcolor{red}{↑19.9\%} & 7.20\% \\
LocAgent & 0.504~\textcolor{red}{↑7.9\%} & 0.630~\textcolor{red}{↑10.5\%} & 0.640~\textcolor{red}{↑15.9\%} & 0.535~\textcolor{red}{↑14.4\%} & 0.566~\textcolor{red}{↑10.2\%} & 0.120~\textcolor{red}{↑135.0\%} & 0.186~\textcolor{red}{↑92.5\%} & 0.196~\textcolor{red}{↑87.8\%} & 0.135~\textcolor{red}{↑110.4\%} & 0.154~\textcolor{red}{↑107.1\%} & 27.20\% \\
\appnameblank & \textbf{0.544}& \textbf{0.696}& \textbf{0.742}& \textbf{0.612}& \textbf{0.624}& \textbf{0.282}& \textbf{0.358}& \textbf{0.368}& \textbf{0.284}& \textbf{0.319}& \textbf{0.00\%}\\
\midrule
\multicolumn{12}{c}{\cellcolor{lb} Qwen2.5-Coder-14B} \\
\midrule
Agentless-FL & 0.578~\textcolor{red}{↑4.2\%} & 0.760~\textcolor{red}{↑0.3\%} & 0.814~\textcolor{red}{↑0.5\%} & 0.637~\textcolor{red}{↑3.1\%} & 0.671~\textcolor{red}{↑2.7\%} & 0.260~\textcolor{red}{↑65.4\%} & 0.336~\textcolor{red}{↑64.3\%} & 0.350~\textcolor{red}{↑68.0\%} & 0.245~\textcolor{red}{↑73.9\%} & 0.301~\textcolor{red}{↑63.8\%} & 3.00\% \\
OrcaLoca & 0.542~\textcolor{red}{↑11.1\%} & 0.574~\textcolor{red}{↑32.8\%} & 0.574~\textcolor{red}{↑42.5\%} & 0.524~\textcolor{red}{↑25.4\%} & 0.557~\textcolor{red}{↑23.7\%} & 0.232~\textcolor{red}{↑85.3\%} & 0.366~\textcolor{red}{↑50.8\%} & 0.378~\textcolor{red}{↑55.6\%} & 0.258~\textcolor{red}{↑65.1\%} & 0.300~\textcolor{red}{↑64.3\%} & 5.40\% \\
LocAgent & 0.556~\textcolor{red}{↑8.3\%} & 0.718~\textcolor{red}{↑6.1\%} & 0.732~\textcolor{red}{↑11.7\%} & 0.607~\textcolor{red}{↑8.2\%} & 0.633~\textcolor{red}{↑8.8\%} & 0.176~\textcolor{red}{↑144.3\%} & 0.370~\textcolor{red}{↑49.2\%} & 0.396~\textcolor{red}{↑48.5\%} & 0.242~\textcolor{red}{↑76.0\%} & 0.271~\textcolor{red}{↑81.9\%} & 14.40\% \\
\appnameblank & \textbf{0.602}& \textbf{0.762}& \textbf{0.818}& \textbf{0.657}& \textbf{0.689}& \textbf{0.430}& \textbf{0.552}& \textbf{0.588}& \textbf{0.426}& \textbf{0.493}& \textbf{0.20\%}\\
\midrule
\multicolumn{12}{c}{\cellcolor{lb} Qwen2.5-Coder-32B} \\
\midrule
Agentless-FL & 0.624~\textcolor{red}{↑2.6\%} & 0.784~\textcolor{red}{↑4.8\%} & 0.834~\textcolor{red}{↑3.6\%} & 0.688~\textcolor{red}{↑2.5\%} & 0.709~\textcolor{red}{↑3.0\%} & 0.326~\textcolor{red}{↑36.8\%} & 0.460~\textcolor{red}{↑31.7\%} & 0.484~\textcolor{red}{↑34.3\%} & 0.339~\textcolor{red}{↑35.4\%} & 0.394~\textcolor{red}{↑34.8\%} & 3.00\% \\
OrcaLoca & 0.612~\textcolor{red}{↑4.6\%} & 0.654~\textcolor{red}{↑25.7\%} & 0.654~\textcolor{red}{↑32.1\%} & 0.590~\textcolor{red}{↑19.5\%} & 0.633~\textcolor{red}{↑15.3\%} & 0.254~\textcolor{red}{↑75.6\%} & 0.468~\textcolor{red}{↑29.5\%} & 0.486~\textcolor{red}{↑33.7\%} & 0.308~\textcolor{red}{↑49.0\%} & 0.359~\textcolor{red}{↑47.9\%} & 4.00\% \\
LocAgent & \textbf{0.686}~\textcolor{green}{↓6.7\%} & 0.806~\textcolor{red}{↑2.0\%} & 0.816~\textcolor{red}{↑5.9\%} & \textbf{0.716}~\textcolor{green}{↓1.5\%} & \textbf{0.743}~\textcolor{green}{↓1.7\%} & 0.250~\textcolor{red}{↑78.4\%} & 0.464~\textcolor{red}{↑30.6\%} & 0.530~\textcolor{red}{↑22.6\%} & 0.325~\textcolor{red}{↑41.2\%} & 0.367~\textcolor{red}{↑44.7\%} & 12.80\% \\
\appnameblank     & 0.640& \textbf{0.822}& \textbf{0.864}& 0.705& 0.730 & \textbf{0.446}& \textbf{0.606}& \textbf{0.650}& \textbf{0.459}& \textbf{0.531}& \textbf{1.00\%}\\
\bottomrule
\end{tabular}
\end{adjustbox}
\label{effect_verified}
\end{table*}

\section{Experiment Result}

\subsection{RQ1: Effectiveness}~\label{sec:rq1}
To validate the effectiveness of \appname in issue localization, we conduct experiments based on the Qwen2.5-coder 7B/14B/32B models. Table~\ref{effect_lite} and Table~\ref{effect_verified} present the experimental results on SWE-bench Lite and SWE-Bench Verified, respectively.

Experimental results show that on SWE-bench Lite, \appname relatively improves file-level Top-1 localization accuracy by 8.28\% on average and function-level Top-1 accuracy by 117.87\%.
On SWE-bench Verified, the average improvements are 4.94\% and 74.21\%, respectively. 
This demonstrates that \appname can localize issues more effectively, especially at the function level, than the baselines.
Besides, although OrcaLoca achieves the highest file-level Top-1 accuracy on SWE-bench Lite using Qwen2.5-coder-7B, and LocAgent achieves the highest file-level Top-1 accuracy on the two benchmarks using Qwen2.5-coder-32B, \appname still achieves the highest Top-5 file-level accuracy and the best function-level performance.
This is because during file-level localization, \appname guides the LLM to explore files beyond those mentioned in the issue description to expand the breadth of the search space, which may lead to a decrease in Top-1 accuracy at file-level.
At the function level, \appname guides the LLM in examining source code to make a deeper search and employs a pruner agent to filter out irrelevant locations to control search direction, which decreases the noise introduced from the file level, resulting in stronger performance. 
Since downstream tasks like issue solving typically rely on finer-grained localization and consider multiple candidate locations, \appname is still able to perform best in downstream tasks.

We also examine the \emph{empty rate} of different methods at the function level.
\appname successfully produces a recommendation list for almost all bugs, while agent-based methods, especially LocAgent, fail to localize by about 20\%. 
We further inspect the frequency of error messages appearing in the context of \appname, and find that the frequency of ``ToolError'' and ``KeyError'' occurrences is significantly reduced.
This is because the reflective alignment of \appname verifies the structure of the model’s output, increasing the success rate of output parsing.

We manually inspect the results to understand \appname's better performance, and summarize two main reasons.
First, \appname expands the breadth of the search space through the module call graph, whereas other baselines are limited to the space defined by the issue description. 
For example, for the case shown in Figure~\ref{motivation example}, Agentless asks LLM to return suspicious files from the issue description, resulting in ``\path{django/forms/fields.py}''.
In contrast, \appname provides the module call graph to the LLM so that the correct module (``\path{django/forms/models.py}'') gets incorporated into the search space at the file level.
Second, \appname performs localization by providing source code within the search space and pruning irrelevant code to control direction, whereas Agentless does not provide source code to the LLM, and OrcaLoca and LocAgent allow the LLM to explore freely without constraints.
For the case shown in Figure~\ref{motivation example}, Agentless prompt LLM by a file skeleton format, which does not contain any implementation source code of functions in the file.
LocAgent allows agents to explore code files themselves, leading to repeatedly calling the ``\emph{traverse\_graph}'' tool to examine the structure of ``\path{django/forms/fields.py}'' twice, and to use the ``\emph{search\_code\_snippets}'' tool to retrieve all context related to ``ForeignKey field'' in files such as ``\path{django/forms/fields.py}'' and ``\path{django/db/models/fields/related.py}'', introducing a large amount of noise.
In contrast, \appname uses a pruner agent to control direction to prevent irrelevant exploration. The pruner filters out unrelated paths (e.g., through ``\emph{RelatedField}''), guiding the search toward the correct function.

\answerRQ{Compared to other baselines, \appname achieves 2.5\%-34.6\% improvement at Top-1 file level and 8.6\%-353.2\% at Top-1 function level with the lowest empty rate of recommendation list, which demonstrates its effectiveness. }

\begin{table*}
    \begin{minipage}{0.7\textwidth}
    \centering
\caption{Ablation study results of \appname on SWE-bench Lite.}
\resizebox{!}{1.3cm}{
\begin{tabular}{lcccccc}
\toprule
\multirow{2}{*}{\qquad\qquad  \textbf{Method}} & \multicolumn{3}{c}{\textbf{File-level}} & \multicolumn{3}{c}{\textbf{Function-level}}  \\
\cmidrule{2-7}
& \textbf{Top-1}   & \textbf{MAP}   & \textbf{MRR}  & \textbf{Top-1}   & \textbf{MAP}   & \textbf{MRR} \\
\midrule
\appnameblank & 0.613& 0.705& 0.701& 0.433& 0.465& 0.493\\
\quad w/o Reflective Alignment  & 0.577 \textcolor{green}{\textdownarrow 5.9\%}& 0.662 \textcolor{green}{\textdownarrow 6.1\%}& 0.660\textcolor{green}{\textdownarrow 5.8\%}& 0.390 \textcolor{green}{\textdownarrow 9.9\%}& 0.424 \textcolor{green}{\textdownarrow 8.8\%}& 0.456 \textcolor{green}{\textdownarrow 7.5\%}\\
\quad w/o Module Call Graph  & 0.540 \textcolor{green}{\textdownarrow 11.9\%}& 0.630 \textcolor{green}{\textdownarrow 10.6\%}& 0.630 \textcolor{green}{\textdownarrow 10.1\%}& 0.383 \textcolor{green}{\textdownarrow 11.5\%}& 0.421 \textcolor{green}{\textdownarrow 9.5\%}& 0.448 \textcolor{green}{\textdownarrow 9.1\%}\\
\quad w/o Iterative Search  & - & - & - & 0.320 \textcolor{green}{\textdownarrow 27.8\%}& 0.372 \textcolor{green}{\textdownarrow 20.0\%}& 0.390 \textcolor{green}{\textdownarrow 20.2\%}\\
\quad w/o Pruning  & - & - & - & 0.413 \textcolor{green}{\textdownarrow 4.6\%}& 0.445 \textcolor{green}{\textdownarrow 4.3\%}& 0.482 \textcolor{green}{\textdownarrow 2.2\%}\\
\bottomrule
\end{tabular}
}
\label{ablation}

    \end{minipage}
    \hfill
    \begin{minipage}{0.29\textwidth}
    \centering
\caption{The impact of iteration on \appname's performance.}
\resizebox{!}{1.3cm}{
\begin{tabular}{cccc}
\toprule
\multirow{2}{*}{\textbf{Method}} & \multicolumn{3}{c}{\textbf{Function-level}}  \\
\cmidrule{2-4}
& \textbf{Top-1}   & \textbf{MAP}   & \textbf{MRR} \\
\midrule
\appnameblank@1  & 0.383 & 0.418 & 0.442  \\
\appnameblank@3  & 0.347 & 0.406 & 0.425  \\
\appnameblank@5  & 0.370 & 0.415 & 0.441  \\
\appnameblank@7  & 0.407 & 0.450 & 0.475 \\
\midrule
\appnameblank@10 & \textbf{0.430} & \textbf{0.461} & \textbf{0.489} \\
\bottomrule
\end{tabular}
}
\label{parmasetting}

    \end{minipage}
\end{table*}

\subsection{RQ2: Ablation}
\appname consists of four key components, i.e., reflective alignment, module call graph, function call graph, and pruning.
To investigate their contributions to \appname's effectiveness, we separately eliminate each of them and evaluate the effectiveness of the obtained variant.
To more clearly observe the performance differences between variants, we select the best-performing setting from RQ1 and conduct this ablation study using Qwen2.5-Coder-32B on SWE-bench Lite. 
We only evaluate the function-level localization performance for iterative search and pruning, as these two components exclusively affect function-level localization.
The results are shown in Table~\ref{ablation}.

\begin{itemize}[left=0pt, topsep=0em]
    \item \textbf{Module Call Graph (Section \ref{step1}):} Eliminating the module call graph essentially removes the file dependency information provided to the LLM, and leads to a 10\% performance loss.
    The reason may be that the LLM is misled by the repository-structure tree and causes the function-level search space to be altered, leading to the incorrect identification of suspicious files. 
    \item \textbf{Iterative Search (Section \ref{step2}):} Eliminating iterative search results in approximately a 20\% performance drop.
    It may deprive \appname of the ability to obtain context from the function call graph, reducing it to make direct judgments based solely on the repository structure tree. Since the repository structure tree lacks concrete code segment information, it no longer provides the full context necessary to support decision-making.
    \item \textbf{Pruning (Section \ref{step3}):} Eliminating pruning contributes approximately 3\% loss to the overall performance.
    It may increase \appname's tendency to explore multiple directions, leading to the inclusion of irrelevant and redundant information in the context. 
    This may introduce noise into the context, making it harder for the LLM to focus on the most relevant code segments.
    \item \textbf{Reflective Alignment (Section \ref{step0}):} Eliminating reflective alignment leads to a performance drop of approximately 6–10\%. This may be because it reduces the LLM's instruction-following capability and self-correction ability, resulting in outputs that cannot be properly parsed and ultimately leading to localization failure.
\end{itemize}

We provide two representative examples, i.e., Django-13315 and Django-12908, to demonstrate the contribution of each component.
\textbf{The detailed description can be found in \appendixurl-C.}
These results indicate that all these components are useful and effective for this task.

Furthermore, we explore the impact of the maximum search iterations (denoted as $maxIter$) on overall performance in the iterative search process. We repeat the function-level localization phase for $maxIter = \{1, 3, 5, 7,10\}$ and record the Top-1, MAP, and MRR metrics. The results are presented in Table \ref{parmasetting}, which shows that as the number of iterations increases, \appname's localization performance first decreases and then increases. This may be because, in \appname, the search process is conducted step by step, meaning the LLM can only access the next adjacent node in the function call graph at each iteration. As a result, the LLM requires a certain number of iterations to gather sufficient context about suspicious nodes. While a small amount of relevant context can aid decision-making to some extent, obtaining the full context of a suspicious node proves to be more effective.

\answerRQ{All the key components of \appname contribute to its localization performance, with iterative search contributing most, accounting for approximately 20\% of the overall performance impact. The number of iterations also influences \appname's effectiveness. Within the range of [1,10], localization performance first decreases and then increases, reaching its peak performance at 10 iterations.}

\subsection{RQ3: Application}\label{guidance}

\begin{table}[t!]
\centering
\caption{The resolution rate of patches generated by different localization methods on SWE-bench.}
\resizebox{\columnwidth}{!}{
\begin{tabular}{cccc|ccc}
\toprule
\multirow{2}{*}{\textbf{Method}} & \multicolumn{3}{c}{\textbf{SWE-bench Lite}}& \multicolumn{3}{c}{\textbf{SWE-bench Verified}}\\
\cmidrule{2-7}
& \textbf{Applied\%}  & \textbf{Resovled\%}  & \textbf{imp.}   & \textbf{Applied\%}  & \textbf{Resovled\%}  &\textbf{imp.}   \\
\midrule
\multicolumn{7}{c}{\cellcolor{lb} Qwen2.5-Coder-32B}\\
\midrule
Agentless & 79.33 & 17.33  & \textcolor{red}{\textuparrow 25.00\%}& 84.80& 27.40&\textcolor{red}{\textuparrow 9.49\%}\\
Orcaloca + Agentless & 73.67& 17.00& \textcolor{red}{\textuparrow 27.47\%}& 71.80& 27.20&\textcolor{red}{\textuparrow 10.29\%}\\
LocAgent + Agentless& 77.33& 19.00& \textcolor{red}{\textuparrow 14.05\%}&79.80 &29.40 &\textcolor{red}{\textuparrow 2.04\%}\\
\appnameblank + Agentless &  \textbf{84.67} & \textbf{21.67}  & -  & \textbf{88.00}& \textbf{30.00}&-\\
\midrule
\multicolumn{7}{c}{\cellcolor{lb} GPT-4o}\\
\midrule
Agentless & 84.67& 22.33& \textcolor{red}{\textuparrow 14.96\%}& 90.80& 32.00&\textcolor{red}{\textuparrow 8.13\%}\\
Orcaloca + Agentless & 79.67 & 19.67& \textcolor{red}{\textuparrow 30.50\%}& 77.40& 31.00&\textcolor{red}{\textuparrow 11.6\%}\\
 LocAgent + Agentless& 81.33& 21.00& \textcolor{red}{\textuparrow 22.24\%}& 82.00& 33.60&\textcolor{red}{\textuparrow 2.98\%}\\
\appnameblank + Agentless &   \textbf{92.00}&   \textbf{25.67}& -  & \textbf{95.40}& \textbf{34.60}& -\\
\bottomrule
\end{tabular}
}
\label{rq2}
\end{table}

To explore whether the improved localization results produced by \appname can help existing software engineering agents improve bug fixing, we use the function-level localization results obtained in RQ1 with Qwen2.5-coder-32B and apply the repair and patch validation phases of Agentless-1.5 for bug fixing.  
We separately generate patches using Qwen2.5-coder-32B and GPT-4o-0806. 
Table \ref{rq2} presents the results.

On SWE-bench Lite, \appnameblank+Agentless increases the resolution rate by 14.05\%–30.5\%, while on SWE-bench Verified, the improvement is 2.04\%–11.6\%. 
This indicates that under the same repair conditions, more accurate localization information not only enhances a model's ability to generate patches that can be successfully applied but also improves the overall bug-fixing performance. 
Compared to Qwen2.5-coder-32B, GPT-4o-0806 demonstrates stronger patch generation capabilities, highlighting the differences in how models inherently understand and utilize localization information.

\begin{figure}[t]
    \centering
    \includegraphics[width=1\linewidth]{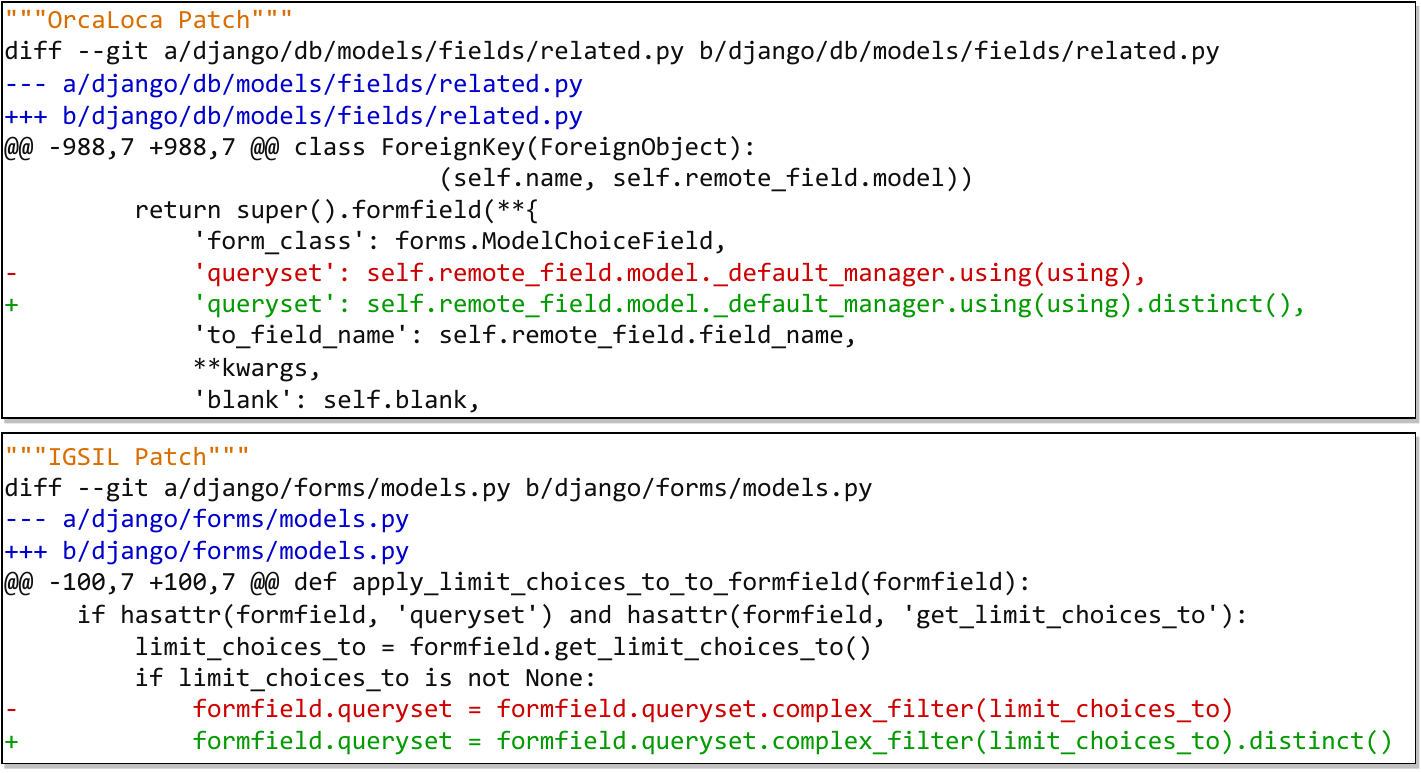}
    \caption{Patch generated by OrcaLoca+Agentless and \appnameblank+Agentless for Django-13315.}
    \label{genpatch}
\end{figure}

Figure~\ref{genpatch} presents an example where incorrect localization information directly leads to incorrect patch generation. In this example, \appnameblank+Agentless and OrcaLoca+Agentless generate different patches for the issue Django-13315, where Agentless fails to generate. Although the LLM correctly identifies that function \texttt{distinct} should be used to address the form duplication problem described in the issue, an incorrect modification location causes the patch to fail the test suite.

\answerRQ{\appname enhances issue solving by providing more accurate localization information. Compared to other baselines, \appnameblank+Agentless achieves a 2.04\%–30.5\% improvement in the resolution rate.}

\subsection{RQ4: Generalizability}\label{generalizability}

\begin{table}[!ht]
\centering
\caption{The localization results and costs of \appname and baselines under different families of LLM.}
\resizebox{\columnwidth}{!}{
\begin{tabular}{cccccc|ccccc}
\toprule
\multirow{2}{*}{\textbf{Method}} & \multicolumn{5}{c}{\textbf{SWE-bench Lite mini}} & \multicolumn{5}{c}{\textbf{SWE-bench Verified mini}} \\
\cmidrule{2-11}
 & \textbf{File@5} & \textbf{Func@1} & \textbf{MAP} & \textbf{Token} & \textbf{\$Cost} & \textbf{File@5} & \textbf{Func@1} & \textbf{MAP} & \textbf{Token} & \textbf{\$Cost} \\
\midrule
\multicolumn{11}{c}{\cellcolor{lb} DeepSeek-v3-0324} \\
\midrule
Agentless-FL & 0.86 & 0.28 &      0.331&  10.0k&  0.003&  \textbf{0.92}&  0.46&     0.494&  10.1k&  0.003\\
OrcaLoca     &      0.76&      0.44&      0.527&  0.12M&  0.026&  0.80&  0.40&     0.507&  0.10M&  0.022\\
LocAgent     & 0.78 & 0.14 &      0.276&  0.53M&  0.055&  0.80&  0.30&     0.390&  0.64M&  0.062\\
\appnameblank        & \textbf{0.90} & \textbf{0.64} &      \textbf{0.659}&  0.07M&  0.022&  \textbf{0.92}&  \textbf{0.66}&     \textbf{0.636}&  0.07M&  0.022\\
\midrule
\multicolumn{11}{c}{\cellcolor{lb} GPT-4o-2024-0806} \\
\midrule
Agentless-FL & \textbf{0.92}& 0.30&       0.453&  10.6k&  0.026&  0.86&  0.58&     0.599&  10.4k&  0.030\\
OrcaLoca     &     0.70&     0.28&       0.387&  0.14M&  0.370&  0.76&  0.36&     0.459&  0.10M&  0.270\\
LocAgent     &     0.76&     0.44&       0.218&  2.07M&  4.004&      0.56&      0.24&     0.290&  2.04M&  3.938\\
\appnameblank        & 0.90& \textbf{0.56}&       \textbf{0.604}&  0.14M&  0.341&  \textbf{0.90}&  \textbf{0.60}&     \textbf{0.620}&  0.12M&  0.304\\
\bottomrule
\end{tabular}

}
\label{general}
\end{table}

To investigate the generalizability of \appname across different families of LLMs, we additionally evaluate the effectiveness of \appname with two state-of-the-art models, i.e., DeepSeek-v3-0324 and GPT-4o-2024-0806, which are representatives of open-source and closed-source models, respectively.
We independently sample 50 instances each from SWE-bench Lite and SWE-bench Verified, referred to as SWE-bench Lite mini and SWE-bench Verified mini, to control experimental costs. 
The instances selected have been released in the replication package.
We report the Top-5 file-level accuracy, Top-1 function-level accuracy, and function-level MAP in Table~\ref{general}. 
\textbf{The remaining metrics are provided in \appendixurl-D}.

On each dataset, \appname achieved the best function-level Top-5 localization accuracy and MAP, which demonstrates its generalization ability across different LLMs.
We further evaluate the execution costs of different methods. 
Although Agentless shows the lowest token consumption and cost, we notice that this method is highly unstable: its function localization performance on SWE-bench Lite mini is only half that of \appname, but when using GPT-4o on SWE-bench Verified mini, it nearly matches \appname.
In addition, we observed that OrcaLoca underperforms compared to \appname, while LocAgent is both less effective and significantly more expensive. 
These results indicate that \appname is the most cost-efficient method apart from Agentless. 

\answerRQ{\appname can generalize across both open-source and closed-source models, and it is also one of the most cost-efficient methods, requiring only \$0.32 per instance with GPT-4o and \$0.02 per instance with DeepSeek-v3.}

\section{Discussion}

\subsection{Evaluating Localization Based on Generated Patches}

Some studies \cite{agentless, orcaloca, ma2024repository} evaluate issue localization performance together with issue solving by examining the percentage that the modification locations in the submitted patch match the corresponding ground truth patch.
They typically use \textit{file match rate} and \textit{function match rate} as the evaluation metrics, which measure the probabilities that the modified file and function align with the ground truth, respectively.
To facilitate comparison, we leverage \appname with GPT-4o-0513 for localization and generate four sets of candidate patches using the repair phase of Agentless-1.5 \cite{agentless} on SWE-bench Lite, following prior studies \cite{agentless,ruan2024specrover,repograph,moatless-tools, infantcoder}.
Each set contains one greedy decoding sample and nine samples with temperature=0.8.
We use the patch validation phase of Agentless-1.5 to select and rerank patches.
We denote this result as \textbf{\appnameblank@40}, and the results in Section \ref{guidance} obtained with Qwen2.5-Coder-32B + GPT-4o are denoted as \appnameblank@10, Agentless@10, Orcaloca@10, and LocAgent@10, as they only generate 10 patches for selection.

We compare \appnameblank@40 with existing issue-solving methods that are also based on GPT-4o-0513, including OpenCSG Starship Agentic Coder \cite{opencsg_starship}, Agentless1.5 \cite{agentless}, AutoCodeRover-v2 \cite{ruan2024specrover}, Infant Coder \cite{infantcoder}, RepoGraph \cite{repograph}, Moatless Tools \cite{moatless-tools}, and SWE-Agent \cite{yang2024sweagent}, in terms of file match rate and function match rate.
Please note that these methods use different pipelines to generate and validate patches.

\begin{table}[t]
\centering
\caption{Performance of methods using GPT-4o on submissions of SWE-bench Lite. * indicates the agent uses GPT-4o-0806 model. \\ \textsuperscript{\dag} indicates the result is reproduced by us.}
\resizebox{0.75\columnwidth}{!}{
\begin{tabular}{cccc}
\toprule
\textbf{Method} & \makecell{\textbf{File}\\\textbf{Matched}}   & \makecell{\textbf{Function}\\\textbf{Matched}}   &\textbf{Resolved\%}\\
\midrule
*OpenCSG Starship \cite{opencsg_starship}& 0.720& 0.523&\textbf{39.7}\\
Agentless-1.5 \cite{agentless}  & 0.697 & 0.517&32.0\\
\textbf{\appnameblank@40}& 0.707& 0.520&31.3\\
AutoCodeRover-v2 \cite{ruan2024specrover}& 0.693& 0.500&30.7\\
Infant-Coder \cite{infantcoder}& \textbf{0.740}& \textbf{0.547}&30.0\\
RepoGraph \cite{repograph}& 0.710& 0.493& 29.7\\
\textsuperscript{\dag}Agentless-1.5 \cite{agentless}  &0.690& 0.480&28.3\\
Moatless-Tools \cite{moatless-tools}& 0.730& 0.507& 24.7\\
SWE-Agent \cite{yang2024sweagent}& 0.580& 0.427&18.3\\
\midrule
\mbox{*}Orcaloca@10& 0.573& 0.413& 19.7\\
\mbox{*}LocAgent@10& 0.603& 0.447& 21.0\\
\mbox{*}Agentless@10& 0.650& 0.450& 22.3\\
\mbox{*}\appnameblank@10& 0.683& 0.483& 25.7\\
\bottomrule
\end{tabular}
}
\label{leaderboard}
\end{table}

We observe that as localization accuracy improves, \appnameblank@10, Agentless@10, OrcaLoca@10, and LocAgent@10, which use the same patch generation and validation pipeline, exhibit a positive correlation between file match rate, function match rate, and resolved\%. 
However, some counterexamples emerge when different patch generation and validation processes are used.
One example is that Infant-Coder, which has the highest function match rate, has a resolved\% that is 9.7\% lower than OpenCSG, which has the second-highest function match rate.
A similar discrepancy occurs between \appnameblank@10 and Moatless-Tools, as shown in Table \ref{leaderboard}.
These results may imply that when using different patch generation and validation pipelines, file and function match rates may not accurately reflect the localization performance.
For example, poor generation and validation methods may fail to select the correct locations and generate unreasonable patches even when correct locations are provided in the candidate locations, leading to a decrease in file and function match rate.

Another interesting result is that \appnameblank@40 performs closely to Agentless1.5, despite \appname demonstrating stronger localization capabilities (c.f. Table~\ref{effect_lite}).
To understand this result, we fully reproduce Agentless-1.5 with its replication package, and conduct a detailed comparison among the intermediate results, including localization results and each set of generated patches, provided by the authors, generated by our reproduced Agentless, and produced by \appnameblank+Agentless, as shown in Table \ref{detailcomparetoagentless}.
Although the reproduced localization performance is close to the official results, the correct patches generated by the official results of Agentless are consistently higher than our reproduced Agentless.
The Top-5 function-level accuracy of \appname is higher than Agentless by 15\%, the resolution rate of \appnameblank+Agentless is slightly lower than Agentless. 
One possible reason behind the performance variations is the randomness caused by high-temperature sampling during patch generation.

\begin{table}[t]
\centering
\caption{A further comparison between \appname and Agentless-1.5 on localization and patch generation phases. \textsuperscript{\dag} indicates the result is reproduced by us. S-n indicates the n-th set of patches.}
\begin{adjustbox}{width=0.8\columnwidth}
\begin{tabular}{ccccccc}
\toprule
\multirow{2}{*}{\textbf{Method}} & \multicolumn{3}{c}{\textbf{File-level}} & \multicolumn{3}{c}{\textbf{Function-level}} \\
\cmidrule{2-7}
& \textbf{Top-1}   & \textbf{Top-3}   & \textbf{Top-5}  & \textbf{Top-1}    & \textbf{Top-3}   & \textbf{Top-5}   \\
\midrule
Agentless-1.5  & 0.630& 0.810& 0.850& 0.250& 0.450& 0.490  \\
\textsuperscript{\dag}Agentless-1.5  & 0.630& 0.817& 0.847& 0.250& 0.407& 0.437  \\
\appnameblank@40 & 0.650& 0.817& 0.883& 0.480& 0.617& 0.640  \\
\midrule
\multicolumn{7}{c}{\textbf{Resolution Rate of Each Set of Patches}} \\
\midrule
& \textbf{S-1}& \textbf{S-2}& \textbf{S-3}  & \textbf{S-4}    & \textbf{All}   & \textbf{Resolved\%}   
\\
Agentless-1.5   & 81/300 & 82/300 & 85/300 & 78/300 & 96/300 & 32.0   
\\
\textsuperscript{\dag}Agentless-1.5   & 72/300 & 72/300 & 70/300 & 70/300 & 86/300 & 28.3   
\\
\appnameblank@40  & 76/300 & 75/300 & 77/300 & 77/300 & 94/300 & 31.3   \\

\bottomrule
\end{tabular} 
\end{adjustbox}
\label{detailcomparetoagentless}
\end{table}

\subsection{Threats to Validity}
\noindent\textbf{Internal Validity.}
\textit{Hyperparameter settings.}
We set a maximum iteration count to control when the search terminates.
Since the performance of \appname varies with different maximum iteration counts, it remains uncertain whether the chosen hyperparameters are optimal.
We have conducted extensive experiments on hyperparameter settings to ensure that the chosen values are optimal within a reasonable range.
\textit{Data Leakage.} 
We conduct experiments using LLMs released after the publication of SWE-bench. Since SWE-bench is constructed from real-world GitHub projects, and GitHub also serves as a source of training data for such models, there is a potential risk of data leakage.
Meanwhile, we observe that the baseline methods also use the latest LLMs, and \appname outperforms them under the same model settings. 
This relative comparison helps mitigate the threat to some extent.

\noindent\textbf{External Validity.}
\textit{Datasets.} 
We evaluate the effectiveness only on SWE-bench Lite and SWE-bench Verified. It remains uncertain whether \appname can generalize to other programming languages or datasets. However, SWE-bench is currently the most widely adopted benchmark. In future work, we plan to assess the effectiveness of \appname using additional benchmarks.

\section{Related Work}

\subsection{LLM-Based Fault Localization}
LLMs have demonstrated remarkable proficiency in localizing the code snippets according to the failure test cases and error messages. 
Wu et al. \cite{wu2023large} demonstrate the applicability of LLMs in fault localization by directly providing erroneous code and logs to ChatGPT, prompting it to identify defective code lines.
Kang et al. \cite{autofl} leverage LLMs' function-calling capabilities and equip them with multiple retrieval tools to enhance their ability to explore code repositories. Xu et al. 
Qin et al. \cite{agentfl} propose AgentFL, which integrates LLMs with static analysis tools to enable repository-level fault localization.
\cite{flexfl} and Wang et al. \cite{wang2024rcagent} combine static analysis and code retrieval tools to construct an LLM Agent that supports decision-making in fault localization.
These methods often take failed test cases and fault information as input, rather than natural language requirement descriptions.

Recent studies \cite{agentless, locagent, orcaloca, chang2025bridging} have also focused on the ability to identify specific code segments based on GitHub issue reports \cite{swebench, swebenchm}. 
Xia et al. \cite{agentless} propose Agentless, which separates issue localization from issue fixing and employs hierarchical localization by directly prompting LLMs. 
Based on this, Chang et al. \cite{chang2025bridging} introduce a method that refines the three levels of localization differently based on file structures, function call chains, and program dependency graphs, achieving multi-granularity localization performance improvements. 
Yu et al. \cite{orcaloca} propose OrcaLoca, which utilizes an LLM-based agent to dynamically schedule attention across different locations and prunes the context based on distance. 
Chen et al. \cite{locagent} propose LocAgent, which integrates multiple repository graphs to construct indexes, enabling search through various tools.

Unlike existing works~\cite{agentless,orcaloca, locagent}, \appname neither expands the search space by building a complete repository index nor restricts it to the issue description. 
Instead, it adopts a two-stage search strategy: at the file level, it broadens the search space using the module call graph; at the function level, it performs an iterative search based on the function call graph, achieving a balance between the breadth and depth of the search space.

\subsection{Agentic Methods for Issue Solving}
Exploring solutions for software issues in real-world code repositories has become a widely studied problem among researchers.
Existing issue-solving methods can be categorized into three main approaches.  
\textit{LLM agent-based methods}, such as SWE-Agent \cite{yang2024sweagent}, AutoCodeRover \cite{zhang2024autocoderover, ruan2024specrover}, Openhands CodeAct \cite{wang2024openhands}, and Moatless Tools \cite{moatless-tools}, equip agents with various tools for accessing and modifying repositories (e.g., bash, LSP), enabling them to autonomously explore and edit code files for bug fixing. 
For example, SWE-Search \cite{antoniades2024swesearch} builds upon this by introducing Monte Carlo Tree Search to explore the solution space for optimal fixes, mitigating the uncertainty caused by LLM sampling.  
\textit{Pipeline-based methods}, represented by Agentless \cite{agentless}, incorporate human prior knowledge by manually identifying and providing predefined contextual information to assist in fixing issues. Methods such as RepoGraph \cite{repograph} and PatchPilot \cite{li2025patchpilot} follow this approach.  
Additionally, researchers explore \textit{fine-tuning open-source models} for issue-solving tasks. SWE-SynInfer \cite{ma2024lingma} and SWE-Gym \cite{pan2024swegym} demonstrate the effectiveness of fine-tuning and establish a paradigm for evaluating fine-tuned models on this task. SWE-Fixer \cite{xie2025swefixer} extends this work by introducing a simpler and more robust method for constructing training data. Furthermore, SWE-RL \cite{wei2025swerl} and SoRFT \cite{ma2025sorft} explore the potential of reinforcement learning for fine-tuning in the patch generation stage following DeepSeek-R1 \cite{guo2025deepseek} and make a great improvement.

\section{Conclusion and Future Work}

We propose \appname, an LLM-driven open-source solution for issue localization without training.
\appname performs dynamic graph construction through prompt-based analysis of target code segments.
It constrains the potential search space using a module call graph and conducts step-by-step iterative search over a function call graph. 
A pruning mechanism is introduced to control the search direction and manage context effectively.
Using reflective alignment to ensure the interaction with the environment is in the correct format.
On SWE-bench Lite, \appname achieves a 65\% success rate in function-level localization.  
In addition, we integrate \appname with other open-source issue-solving methods, resulting in a 2.04\%–30.5\% improvement in resolution rate. 
This demonstrates that more accurate localization results in a higher issue resolution rate.
In the future, we plan to extend \appname to support more programming languages and evaluation benchmarks. 
We also aim to enhance issue-solving agents' ability to utilize localization results more effectively.

\section*{Acknowledgment}
This research/project was partially supported by the National Natural Science Foundation of China (No. 62202420, No.62302437), and Zhejiang Provincial Natural Science Foundation of China (No.LZ25F020003).

\balance
\bibliographystyle{IEEEtran}
\bibliography{references}

\begin{thebibliography}{10}
\providecommand{\url}[1]{#1}
\csname url@samestyle\endcsname
\providecommand{\newblock}{\relax}
\providecommand{\bibinfo}[2]{#2}
\providecommand{\BIBentrySTDinterwordspacing}{\spaceskip=0pt\relax}
\providecommand{\BIBentryALTinterwordstretchfactor}{4}
\providecommand{\BIBentryALTinterwordspacing}{\spaceskip=\fontdimen2\font plus
\BIBentryALTinterwordstretchfactor\fontdimen3\font minus
  \fontdimen4\font\relax}
\providecommand{\BIBforeignlanguage}[2]{{%
\expandafter\ifx\csname l@#1\endcsname\relax
\typeout{** WARNING: IEEEtran.bst: No hyphenation pattern has been}%
\typeout{** loaded for the language `#1'. Using the pattern for}%
\typeout{** the default language instead.}%
\else
\language=\csname l@#1\endcsname
\fi
#2}}
\providecommand{\BIBdecl}{\relax}
\BIBdecl

\bibitem{swebench}
C.~E. Jimenez, J.~Yang, A.~Wettig, S.~Yao, K.~Pei, O.~Press, and K.~Narasimhan,
  ``Swe-bench: Can language models resolve real-world github issues?''
  \emph{arXiv preprint arXiv:2310.06770}, 2023.

\bibitem{swebenchm}
J.~Yang, C.~E. Jimenez, A.~L. Zhang, K.~Lieret, J.~Yang, X.~Wu, O.~Press,
  N.~Muennighoff, G.~Synnaeve, K.~R. Narasimhan \emph{et~al.}, ``Swe-bench
  multimodal: Do ai systems generalize to visual software domains?''
  \emph{arXiv preprint arXiv:2410.03859}, 2024.

\bibitem{zhao2024enhancing}
J.~Zhao, D.~Yang, L.~Zhang, X.~Lian, Z.~Yang, and F.~Liu, ``Enhancing automated
  program repair with solution design,'' in \emph{Proceedings of the 39th
  IEEE/ACM International Conference on Automated Software Engineering}, 2024,
  pp. 1706--1718.

\bibitem{yin2024thinkrepair}
X.~Yin, C.~Ni, S.~Wang, Z.~Li, L.~Zeng, and X.~Yang, ``Thinkrepair:
  Self-directed automated program repair,'' in \emph{Proceedings of the 33rd
  ACM SIGSOFT International Symposium on Software Testing and Analysis}, 2024,
  pp. 1274--1286.

\bibitem{bouzenia2025repairagent}
I.~Bouzenia, P.~Devanbu, and M.~Pradel, ``Repairagent: An autonomous, llm-based
  agent for program repair,'' in \emph{2025 IEEE/ACM 47th International
  Conference on Software Engineering (ICSE)}.\hskip 1em plus 0.5em minus
  0.4em\relax IEEE Computer Society, 2025, pp. 694--694.

\bibitem{li2025feabenchbenchmarkevaluatingrepositorylevel}
\BIBentryALTinterwordspacing
W.~Li, X.~Zhang, Z.~Guo, S.~Mao, W.~Luo, G.~Peng, Y.~Huang, H.~Wang, and S.~Li,
  ``Fea-bench: A benchmark for evaluating repository-level code generation for
  feature implementation,'' 2025. [Online]. Available:
  \url{https://arxiv.org/abs/2503.06680}
\BIBentrySTDinterwordspacing

\bibitem{deng2025nocode}
L.~Deng, Z.~Jiang, J.~Cao, M.~Pradel, and Z.~Liu, ``Nocode-bench: A benchmark
  for evaluating natural language-driven feature addition,'' \emph{arXiv
  preprint arXiv:2507.18130}, 2025.

\bibitem{meng2024empiricalstudyllmbasedagents}
\BIBentryALTinterwordspacing
X.~Meng, Z.~Ma, P.~Gao, and C.~Peng, ``An empirical study on llm-based agents
  for automated bug fixing,'' 2024. [Online]. Available:
  \url{https://arxiv.org/abs/2411.10213}
\BIBentrySTDinterwordspacing

\bibitem{agentless}
C.~S. Xia, Y.~Deng, S.~Dunn, and L.~Zhang, ``Agentless: Demystifying llm-based
  software engineering agents,'' \emph{arXiv preprint arXiv:2407.01489}, 2024.

\bibitem{orcaloca}
Z.~Yu, H.~Zhang, Y.~Zhao, H.~Huang, M.~Yao, K.~Ding, and J.~Zhao, ``Orcaloca:
  An llm agent framework for software issue localization,'' \emph{arXiv
  preprint arXiv:2502.00350}, 2025.

\bibitem{locagent}
\BIBentryALTinterwordspacing
Z.~Chen, X.~Tang, G.~Deng, F.~Wu, J.~Wu, Z.~Jiang, V.~Prasanna, A.~Cohan, and
  X.~Wang, ``Locagent: Graph-guided llm agents for code localization,'' 2025.
  [Online]. Available: \url{https://arxiv.org/abs/2503.09089}
\BIBentrySTDinterwordspacing

\bibitem{chang2025bridging}
J.~Chang, X.~Zhou, L.~Wang, D.~Lo, and B.~Li, ``Bridging bug localization and
  issue fixing: A hierarchical localization framework leveraging large language
  models,'' \emph{arXiv preprint arXiv:2502.15292}, 2025.

\bibitem{repograph}
S.~Ouyang, W.~Yu, K.~Ma, Z.~Xiao, Z.~Zhang, M.~Jia, J.~Han, H.~Zhang, and
  D.~Yu, ``Repograph: Enhancing ai software engineering with repository-level
  code graph,'' \emph{arXiv preprint arXiv:2410.14684}, 2024.

\bibitem{chen2025unveiling}
Z.~Chen, W.~Ma, and L.~Jiang, ``Unveiling pitfalls: Understanding why ai-driven
  code agents fail at github issue resolution,'' \emph{arXiv preprint
  arXiv:2503.12374}, 2025.

\bibitem{wang2024openhands}
X.~Wang, B.~Li, Y.~Song, F.~F. Xu, X.~Tang, M.~Zhuge, J.~Pan, Y.~Song, B.~Li,
  J.~Singh \emph{et~al.}, ``Openhands: An open platform for ai software
  developers as generalist agents,'' in \emph{The Thirteenth International
  Conference on Learning Representations}, 2024.

\bibitem{liu2024large}
J.~Liu, K.~Wang, Y.~Chen, X.~Peng, Z.~Chen, L.~Zhang, and Y.~Lou, ``Large
  language model-based agents for software engineering: A survey,'' \emph{arXiv
  preprint arXiv:2409.02977}, 2024.

\bibitem{yang2024sweagent}
J.~Yang, C.~Jimenez, A.~Wettig, K.~Lieret, S.~Yao, K.~Narasimhan, and O.~Press,
  ``Swe-agent: Agent-computer interfaces enable automated software
  engineering,'' \emph{Advances in Neural Information Processing Systems},
  vol.~37, pp. 50\,528--50\,652, 2024.

\bibitem{bai2024longalignrecipelongcontext}
\BIBentryALTinterwordspacing
Y.~Bai, X.~Lv, J.~Zhang, Y.~He, J.~Qi, L.~Hou, J.~Tang, Y.~Dong, and J.~Li,
  ``Longalign: A recipe for long context alignment of large language models,''
  2024. [Online]. Available: \url{https://arxiv.org/abs/2401.18058}
\BIBentrySTDinterwordspacing

\bibitem{shinn2023reflexionlanguageagentsverbal}
\BIBentryALTinterwordspacing
N.~Shinn, F.~Cassano, E.~Berman, A.~Gopinath, K.~Narasimhan, and S.~Yao,
  ``Reflexion: Language agents with verbal reinforcement learning,'' 2023.
  [Online]. Available: \url{https://arxiv.org/abs/2303.11366}
\BIBentrySTDinterwordspacing

\bibitem{hui2024qwen2}
B.~Hui, J.~Yang, Z.~Cui, J.~Yang, D.~Liu, L.~Zhang, T.~Liu, J.~Zhang, B.~Yu,
  K.~Lu \emph{et~al.}, ``Qwen2. 5-coder technical report,'' \emph{arXiv
  preprint arXiv:2409.12186}, 2024.

\bibitem{swebenchverified}
\BIBentryALTinterwordspacing
OpenAI, ``Introducing swe-bench verified,'' 2024, accessed: 2025-03-16.
  [Online]. Available:
  \url{https://openai.com/index/introducing-swe-bench-verified/}
\BIBentrySTDinterwordspacing

\bibitem{guo2025deepseek}
D.~Guo, D.~Yang, H.~Zhang, J.~Song, R.~Zhang, R.~Xu, Q.~Zhu, S.~Ma, P.~Wang,
  X.~Bi \emph{et~al.}, ``Deepseek-r1: Incentivizing reasoning capability in
  llms via reinforcement learning,'' \emph{arXiv preprint arXiv:2501.12948},
  2025.

\bibitem{openai2024gpt4ocard}
\BIBentryALTinterwordspacing
OpenAI, ``Gpt-4o system card,'' 2024. [Online]. Available:
  \url{https://arxiv.org/abs/2410.21276}
\BIBentrySTDinterwordspacing

\bibitem{zhang2025sealignalignmenttrainingsoftware}
\BIBentryALTinterwordspacing
K.~Zhang, H.~Zhang, G.~Li, J.~You, J.~Li, Y.~Zhao, and Z.~Jin, ``Sealign:
  Alignment training for software engineering agent,'' 2025. [Online].
  Available: \url{https://arxiv.org/abs/2503.18455}
\BIBentrySTDinterwordspacing

\bibitem{zhang2024autocoderover}
Y.~Zhang, H.~Ruan, Z.~Fan, and A.~Roychoudhury, ``Autocoderover: Autonomous
  program improvement,'' in \emph{Proceedings of the 33rd ACM SIGSOFT
  International Symposium on Software Testing and Analysis}, 2024, pp.
  1592--1604.

\bibitem{flexfl}
C.~Xu, Z.~Liu, X.~Ren, G.~Zhang, M.~Liang, and D.~Lo, ``Flexfl: Flexible and
  effective fault localization with open-source large language models,''
  \emph{arXiv preprint arXiv:2411.10714}, 2024.

\bibitem{autofl}
S.~Kang, G.~An, and S.~Yoo, ``A quantitative and qualitative evaluation of
  llm-based explainable fault localization,'' \emph{Proceedings of the ACM on
  Software Engineering}, vol.~1, no. FSE, pp. 1424--1446, 2024.

\bibitem{agentfl}
Y.~Qin, S.~Wang, Y.~Lou, J.~Dong, K.~Wang, X.~Li, and X.~Mao, ``Soapfl: A
  standard operating procedure for llm-based method-level fault localization,''
  \emph{IEEE Transactions on Software Engineering}, vol.~51, no.~4, pp.
  1173--1187, 2025.

\bibitem{kochhar2016practitioners}
P.~S. Kochhar, X.~Xia, D.~Lo, and S.~Li, ``Practitioners' expectations on
  automated fault localization,'' in \emph{Proceedings of the 25th
  international symposium on software testing and analysis}, 2016, pp.
  165--176.

\bibitem{vllm2024}
\BIBentryALTinterwordspacing
vLLM Team, ``vllm documentation,'' 2024, accessed: 2025-03-16. [Online].
  Available: \url{https://docs.vllm.ai/en/stable/}
\BIBentrySTDinterwordspacing

\bibitem{ma2024repository}
Z.~Ma, S.~An, Z.~Lin, Y.~Zou, and B.~Xie, ``Repository structure-aware training
  makes slms better issue resolver,'' \emph{arXiv preprint arXiv:2412.19031},
  2024.

\bibitem{ruan2024specrover}
H.~Ruan, Y.~Zhang, and A.~Roychoudhury, ``Specrover: Code intent extraction via
  llms,'' \emph{arXiv preprint arXiv:2408.02232}, 2024.

\bibitem{moatless-tools}
Aorwall, ``Moatless tools,'' \url{https://github.com/aorwall/moatless-tools},
  2025, accessed: 2025-03-16.

\bibitem{infantcoder}
B.~Lei, Y.~Li, Y.~Zeng, T.~Ren, Y.~Luo, T.~Shi, Z.~Gao, Z.~Hu, W.~Kang, and
  Q.~Chen, ``Infant agent: A tool-integrated, logic-driven agent with
  cost-effective api usage,'' \emph{arXiv preprint arXiv:2411.01114}, 2024.

\bibitem{opencsg_starship}
\BIBentryALTinterwordspacing
{OpenCSG Team}, ``Opencsg starship,'' 2025, accessed: 2025-03-19. [Online].
  Available: \url{https://opencsg.com/starship}
\BIBentrySTDinterwordspacing

\bibitem{wu2023large}
Y.~Wu, Z.~Li, J.~M. Zhang, M.~Papadakis, M.~Harman, and Y.~Liu, ``Large
  language models in fault localisation,'' \emph{arXiv preprint
  arXiv:2308.15276}, 2023.

\bibitem{wang2024rcagent}
Z.~Wang, Z.~Liu, Y.~Zhang, A.~Zhong, J.~Wang, F.~Yin, L.~Fan, L.~Wu, and
  Q.~Wen, ``Rcagent: Cloud root cause analysis by autonomous agents with
  tool-augmented large language models,'' in \emph{Proceedings of the 33rd ACM
  International Conference on Information and Knowledge Management}, 2024, pp.
  4966--4974.

\bibitem{antoniades2024swesearch}
A.~Antoniades, A.~{\"O}rwall, K.~Zhang, Y.~Xie, A.~Goyal, and W.~Wang,
  ``Swe-search: Enhancing software agents with monte carlo tree search and
  iterative refinement,'' \emph{arXiv preprint arXiv:2410.20285}, 2024.

\bibitem{li2025patchpilot}
H.~Li, Y.~Tang, S.~Wang, and W.~Guo, ``Patchpilot: A stable and cost-efficient
  agentic patching framework,'' \emph{arXiv preprint arXiv:2502.02747}, 2025.

\bibitem{ma2024lingma}
Y.~Ma, R.~Cao, Y.~Cao, Y.~Zhang, J.~Chen, Y.~Liu, Y.~Liu, B.~Li, F.~Huang, and
  Y.~Li, ``Lingma swe-gpt: An open development-process-centric language model
  for automated software improvement,'' \emph{arXiv preprint arXiv:2411.00622},
  2024.

\bibitem{pan2024swegym}
J.~Pan, X.~Wang, G.~Neubig, N.~Jaitly, H.~Ji, A.~Suhr, and Y.~Zhang, ``Training
  software engineering agents and verifiers with swe-gym,'' \emph{arXiv
  preprint arXiv:2412.21139}, 2024.

\bibitem{xie2025swefixer}
C.~Xie, B.~Li, C.~Gao, H.~Du, W.~Lam, D.~Zou, and K.~Chen, ``Swe-fixer:
  Training open-source llms for effective and efficient github issue
  resolution,'' \emph{arXiv preprint arXiv:2501.05040}, 2025.

\bibitem{wei2025swerl}
Y.~Wei, O.~Duchenne, J.~Copet, Q.~Carbonneaux, L.~Zhang, D.~Fried, G.~Synnaeve,
  R.~Singh, and S.~I. Wang, ``Swe-rl: Advancing llm reasoning via reinforcement
  learning on open software evolution,'' \emph{arXiv preprint
  arXiv:2502.18449}, 2025.

\bibitem{ma2025sorft}
Z.~Ma, C.~Peng, P.~Gao, X.~Meng, Y.~Zou, and B.~Xie, ``Sorft: Issue resolving
  with subtask-oriented reinforced fine-tuning,'' \emph{arXiv preprint
  arXiv:2502.20127}, 2025.

\end{thebibliography}

\newpage

\twocolumn[
\begin{center}
    {\Huge Issue Localization via LLM-Driven\\[1ex] Iterative Code Graph Searching} \\
    \vspace{1em}
    {\LARGE Supplementary Material}
\end{center}
\vspace{1em}
]

\section*{A. Call Graph in Code Repository}

\begin{figure}[h!]
    \centering
    \includegraphics[width=\linewidth]{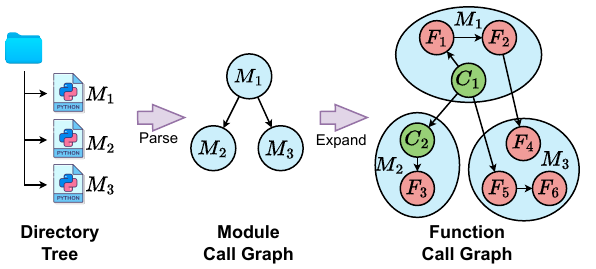}
    \caption{An example of call graph construction.}
    \label{example}
\end{figure}

As shown in Figure \ref{example}, in the root directory, module \( M_1 \) imports modules \( M_2 \) and \( M_3 \), and their invocation relationships are represented in the module call graph. Furthermore, \( M_1 \) contains class \( C_1 \) and functions \( F_1 \) and \( F_2 \), allowing the nodes in the module call graph to be expanded, as illustrated in the function call graph on the right.

We further provided the algorithm used by \appname to construct the call graph using the LLM.

\begin{algorithm}
\caption{Call Graph Construction}
\begin{algorithmic}[1]
    \Statex \textbf{Input:} $filePath$ \Comment{The file path of target entity}
    \Statex \textbf{Input:} $entityName$ \Comment{The target entity name}
    \Statex \textbf{Output:} $CallGraph$ 
    
    \State code = extractCode(filePath, entityName)
    \State importStmts = extractImports(code) \Comment{Parse import statements}
    \State prompt = constructPrompt(importStmts, code)
    \State response = callLLM(prompt)  \Comment{Instruct LLM to analyze the call relations and output it in textual format}
    \State callGraph = parseGraphFromResponse(response)
    \State \Return callGraph
\end{algorithmic}
\label{alg:function-call-graph}
\end{algorithm}

\section*{B. Function Call Graph Searching Algorithm}

We provided the Algorithm \ref{AL1} described in Section \ref{step2}.

\begin{algorithm}
\caption{Function Call Graph Searching with Pruning}
\begin{algorithmic}[1]
    \Statex \textbf{Input:} $\mathcal{G}_F = (\mathcal{V}_f, \mathcal{E}_f)$ \Comment{Function call graph}
    \Statex \textbf{Input:} maxIter \Comment{Maximum of iterations}
    \Statex \textbf{Output:} contextNodes \Comment{The relevant functions or classes}
    
    \State contextNodes=\{\};
    \State visableNodes=initStartPoints($\mathcal{V}_f$);   \Comment{Pre-select by LLM}
    \State counter=0;
    \While{counter$<$maxIter} \Comment{Start Searching}
    \State targetNode=Seacher(visableNodes); \Comment{Search Step}
    \If{targetNode == $null$}
        \State \textbf{break}
    \EndIf
    \State nodeContext=getCode(targetNode);
    \State pruningFlag = Pruner(nodeContext);
        \If{pruningFlag} \Comment{Pruning context}
        \State visableNodes = visableNodes - targetNode ;
        \Else
        \State nextNeighbors = getNeighbors(targetNode) ;
        \State visableNodes = visableNodes $\cup$ nextNeighbors ;
        \State contextNodes = contextNodes $\cup$ targetNode ;
        \EndIf
    \State counter=counter+1;
    \EndWhile
    \State \Return contextNodes;
\end{algorithmic}
\label{AL1}
\end{algorithm}

\section*{C. Examples Demonstrating the Effectiveness of Key Components in \appname}

To better understand the performance differences among the four variants in the ablation study, we manually inspect the experimental results and select two highly representative examples.

\begin{figure}[h]
    \centering
    \includegraphics[width=1\linewidth]{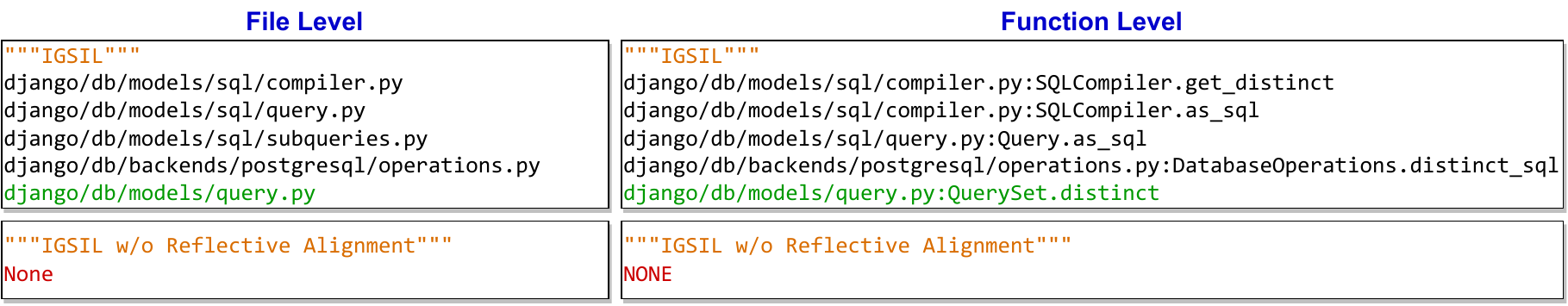}
    \caption{An example of localization failure after removing reflective alignment from Django-12908.}
    \label{ablation_eg2}
\end{figure}

\begin{figure*}[htbp]
    \centering
    \includegraphics[width=1\linewidth]{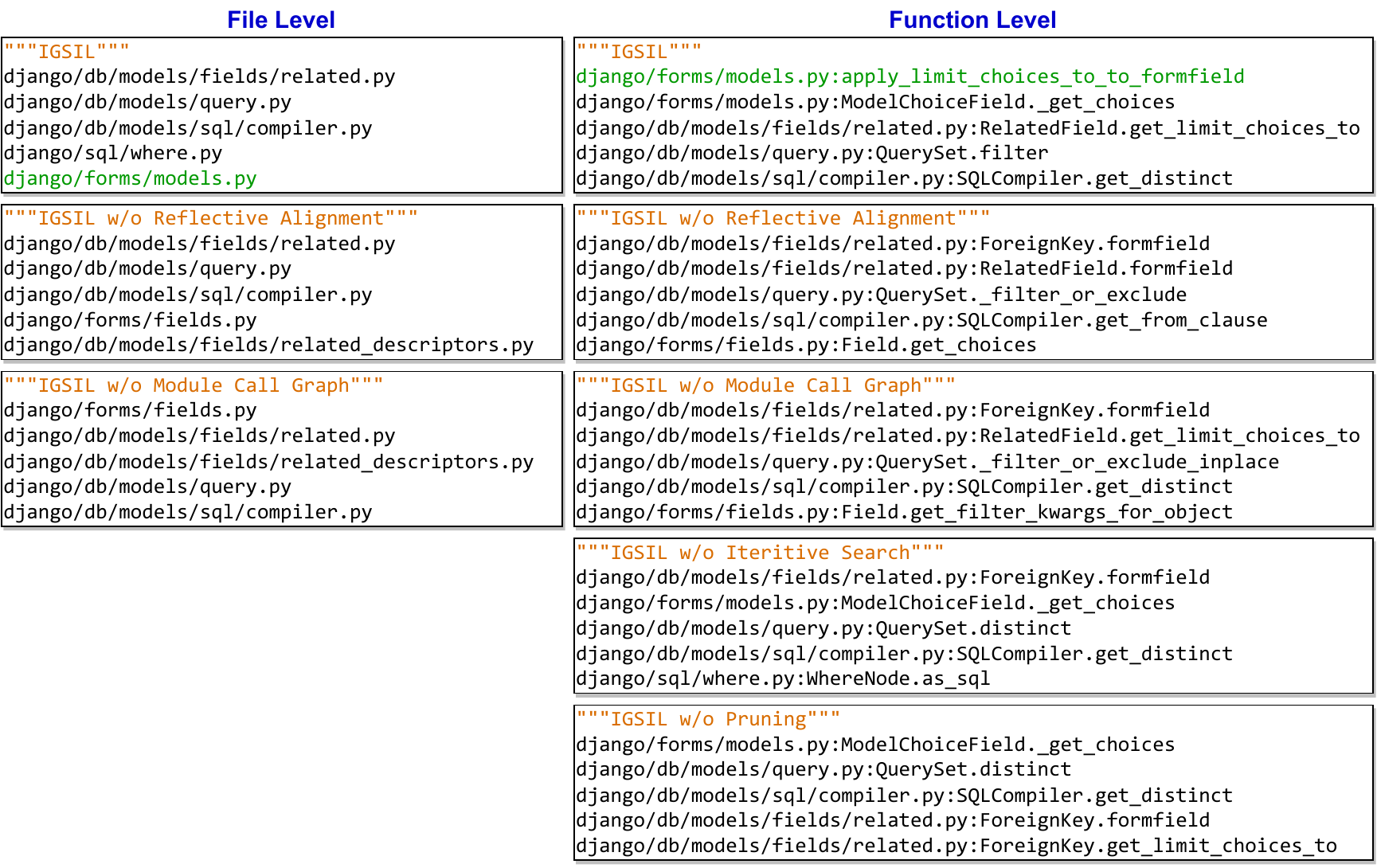}
    \caption{An example where the removal of any core component leads to localization failure from Django-13315.}
    \label{ablation_eg1}
\end{figure*}

We find that without reflective alignment, \appname not only triggers more ``\emph{KeyErrors}'', making it impossible to parse the localization results, but also loses the potential for result reordering and error correction.
As shown in Figure \ref{ablation_eg2}, for the case of Django-12908, all output paths produced by \appname without reflective alignment are incomplete, containing only the final file name without the full path starting from the root directory ``\path{django/}'', making it impossible for the system to parse them.
As shown in Figure \ref{ablation_eg1}, for Django-13315, \appname without reflective alignment fails to reorder and correct the retrieved files, causing the target file ``\path{django/forms/models.py}'' to be excluded from the top five candidate files, thus directly losing the correct search space.

Without the module call graph, \appname exhibits an overly narrow search space at the file level. 
As shown in Figure \ref{ablation_eg1}, the LLM has no way of obtaining information about the target file ``\path{django/forms/models.py}'' from any source, thereby restricting the search space to the issue description and ultimately causing subsequent localization failure.

Without iterative search, although the correct search space is identified, the function signature information within the file skeleton is insufficient to support the LLM in making accurate judgments. 
This approach loses the depth of search, forcing the LLM to match function signatures based solely on the issue description, leading to the selection of ``\path{django/db/models/fields/related.py: ForeignKey.formfield}''.

Without pruning, we observed that the LLM tends to conduct searches in multiple directions. It sequentially queries almost all class nodes in ``\path{django/db/models/fields/related.py}'' and ``\path{django/forms/models.py}''. This not only consumes a large portion of the context window but also exhausts the number of allowed iterations, causing the algorithm to terminate before it reaches the exploration of ``\emph{apply\_limit\_choices\_to\_to\_formfield}''.

\section*{D. RQ4's Results on Different LLMs.}

Table~\ref{full_rq4} shows the complete results of experiments run on different families of LLMs.
\appname outperforms all baseline methods at the function level, indicating that \appname generalizes well to different families of LLMs.
In addition, we observed that Agentless-FL demonstrates the strongest Top-1 file-level localization, which may be because \appname expands the search space at the file level, resulting in some additional modules being included in the localization results.
Agentless-FL’s results on SWE-bench Verified mini are nearly on par with \appname, especially when using GPT-4o. This may be because the issue descriptions in SWE-bench Verified are of higher quality and contain information about the target function of the task.

\begin{table}[t!]
\centering
\caption{Localization results under different families of LLMs on SWE-bench Lite mini and Verified mini.}
\begin{adjustbox}{width=\columnwidth}
\begin{tabular}{cccccc|ccccc|c}
\toprule
\multicolumn{12}{c}{ \textbf{SWE-bench Lite mini} } \\
\midrule
\multirow{2}{*}{\textbf{Method}} & \multicolumn{5}{c}{\textbf{File-level}}  & \multicolumn{5}{c}{\textbf{Function-level}} &  \\
\cmidrule{2-12}
& \textbf{Top-1}   & \textbf{Top-3}   & \textbf{Top-5} & \textbf{MAP} & \textbf{MRR}  & \textbf{Top-1}   & \textbf{Top-3}   & \textbf{Top-5}  & \textbf{MAP} & \textbf{MRR} & \textbf{ER} \\
\midrule
\multicolumn{12}{c}{\cellcolor{lb} Deepseek-v3-0324} \\
\midrule
Agentless-FL & \textbf{0.78}& \textbf{0.84}& 0.86& 0.811& 0.811& 0.28& 0.42& 0.42& 0.331& 0.339& \textbf{0.00}\\
OrcaLoca & 0.76& 0.76& 0.76& 0.76& 0.76& 0.44& 0.64& 0.64& 0.527& 0.540& 0.08\\
LocAgent & 0.72& 0.78& 0.78& 0.747& 0.747& 0.14& 0.40& 0.54& 0.276& 0.292& 0.16\\
\appnameblank & \textbf{0.78}& \textbf{0.84}& \textbf{0.90}& \textbf{0.813}& \textbf{0.813}& \textbf{0.64}& \textbf{0.67}& \textbf{0.78}& \textbf{0.659}& \textbf{0.697}& \textbf{0.00}\\
\midrule
\multicolumn{12}{c}{\cellcolor{lb} GPT-4o-2024-0806} \\
\midrule
Agentless-FL & \textbf{0.76}& 0.86& \textbf{0.92}& \textbf{0.816}& \textbf{0.816}& 0.30& 0.66& 0.70& 0.453& 0.471& \textbf{0.00}\\
OrcaLoca & 0.66& 0.70& 0.70& 0.680& 0.680& 0.28& 0.50& 0.58& 0.387& 0.392& 0.02\\
LocAgent & 0.62& 0.76& 0.76& 0.683& 0.683& 0.10& 0.32& 0.44& 0.218& 0.224& 0.16\\
\appnameblank & 0.70& \textbf{0.88}& \textbf{0.92}& 0.785& 0.785& \textbf{0.56}& \textbf{0.70}& \textbf{0.74}& \textbf{0.604}& \textbf{0.630}& \textbf{0.00}\\
\bottomrule
\toprule
\multicolumn{12}{c}{ \textbf{SWE-bench Verified mini} } \\
\midrule
\multirow{2}{*}{\textbf{Method}} & \multicolumn{5}{c}{\textbf{File-level}}  & \multicolumn{5}{c}{\textbf{Function-level}} &  \\
\cmidrule{2-12}
& \textbf{Top-1}   & \textbf{Top-3}   & \textbf{Top-5} & \textbf{MAP} & \textbf{MRR}  & \textbf{Top-1}   & \textbf{Top-3}   & \textbf{Top-5}  & \textbf{MAP} & \textbf{MRR} & \textbf{ER} \\
\midrule
\multicolumn{12}{c}{\cellcolor{lb} Deepseek-v3-0324} \\
\midrule
Agentless-FL & \textbf{0.80}& 0.86& \textbf{0.92}& \textbf{0.818}& 0.836& 0.46& 0.64& 0.64& 0.494& 0.549& \textbf{0.00}\\
OrcaLoca & 0.74& 0.80& 0.80& 0.740& 0.770& 0.40& 0.76& 0.76& 0.507& 0.507& 0.04\\
LocAgent & 0.68& 0.78& 0.80& 0.706& 0.732& 0.30& 0.54& 0.60& 0.390& 0.524& 0.14\\
\appnameblank & 0.78& \textbf{0.92}& \textbf{0.92}& 0.815& \textbf{0.847}& \textbf{0.66}& \textbf{0.76}& \textbf{0.76}& \textbf{0.636}& \textbf{0.703}& \textbf{0.00}\\
\midrule
\multicolumn{12}{c}{\cellcolor{lb} GPT-4o-2024-0806} \\
\midrule
Agentless-FL & \textbf{0.76}& 0.84& 0.86& 0.782& 0.802& 0.58& \textbf{0.76}& \textbf{0.78}& 0.599& 0.669& \textbf{0.00}\\
OrcaLoca & 0.70& 0.74& 0.76& 0.700& 0.725& 0.36& 0.60& 0.62& 0.459& 0.472& 0.04\\
LocAgent & 0.50& 0.56& 0.56& 0.517& 0.527& 0.24& 0.38& 0.46& 0.290& 0.324& 0.40\\
\appnameblank & 0.74& \textbf{0.88}& \textbf{0.90}& \textbf{0.791}& \textbf{0.807}& \textbf{0.60}& \textbf{0.76}& \textbf{0.78}& \textbf{0.620}& \textbf{0.672}& \textbf{0.00}\\
\bottomrule
\end{tabular}
\end{adjustbox}
\label{full_rq4}
\end{table}

\section*{E. Prompt Templates}
\clearpage
The prompts used in \appname are shown as follows.
\begin{figure}[htbp]
    \centering
    \includegraphics[width=1\columnwidth]{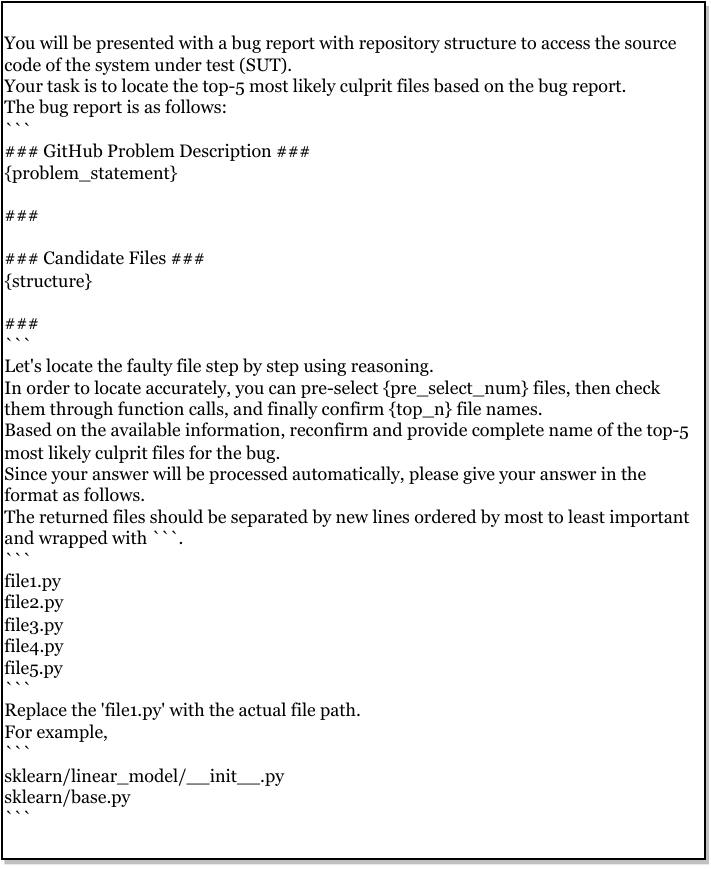}
    \caption{Prompt for Pre-Selection Code Files.}
\end{figure}

\begin{figure}[htbp]
    \centering
    \includegraphics[width=1\linewidth]{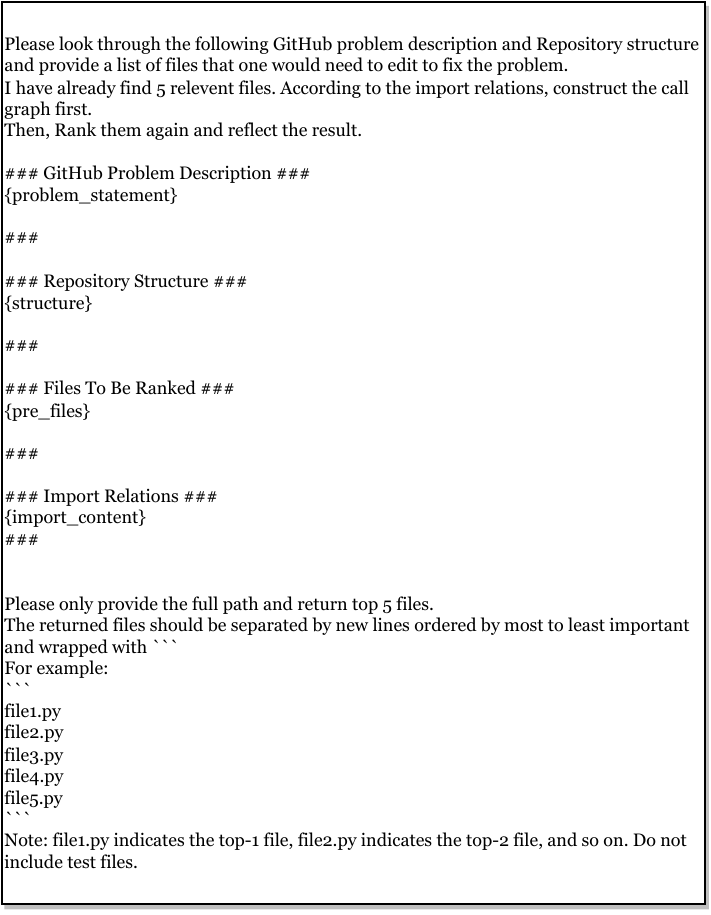}
    \caption{Prompt for Module Call Graph Enhanced Search Space Reduction.}
\end{figure}
\begin{figure}[htbp]
    \centering
    \includegraphics[width=1\linewidth]{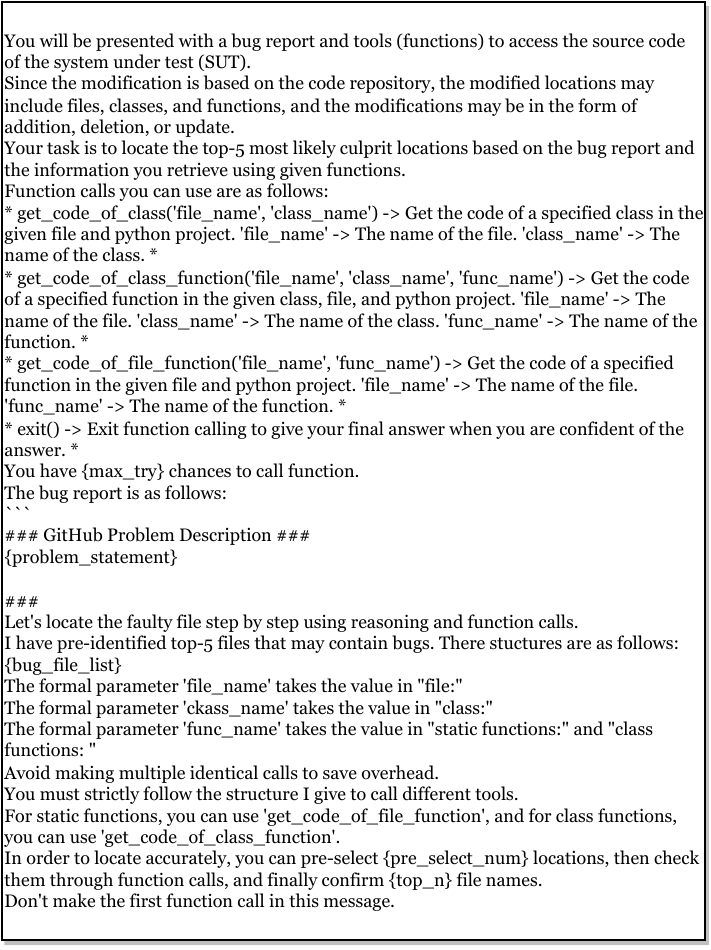}
    \caption{Prompt for Iterative Search.}
\end{figure}

\begin{figure}[htbp]
    \centering
    \includegraphics[width=1\linewidth]{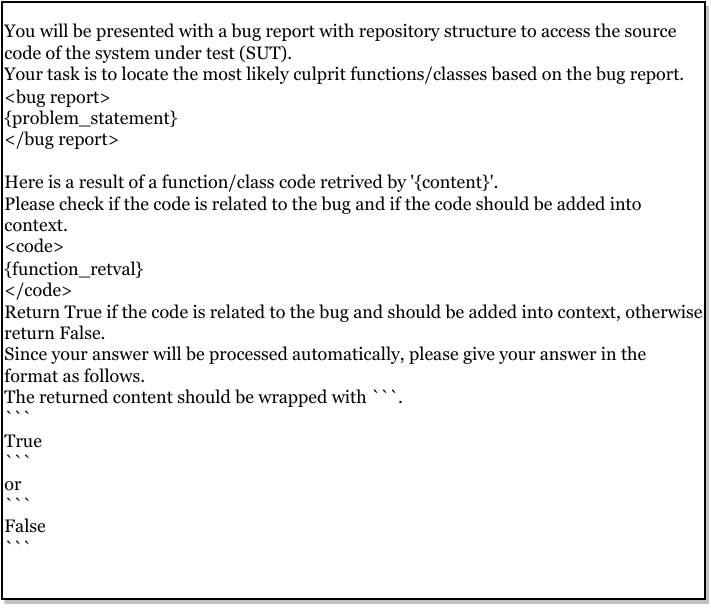}
    \caption{Prompt for Pruning.}
\end{figure}
\begin{figure}[htbp]
    \centering
    \includegraphics[width=1\linewidth]{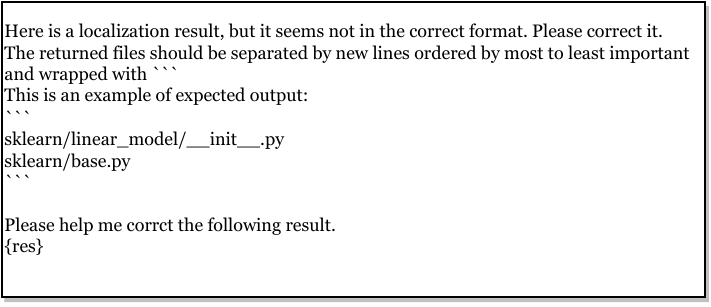}
    \caption{Prompt for Reflective Alignment}
\end{figure}
\clearpage

\end{document}